# Unveiling SN 2022eyw: A Bright Member of the Type Iax Supernova Subclass

Hrishav Das[1,2], Devendra K. Sahu[1], Anirban Dutta[3], Mridweeka Singh[1], G. C. Anupama[1], and Rishabh Singh Teja[1,4]

[1] Indian Institute of Astrophysics, Koramangala 2nd Block, Bangalore 560034, India; hrishav.das@iiap.res.in
[2] Pondicherry University, R.V. Nagar, Kalapet, 605014, Puducherry, India
[3] Department of Physics and Astronomy, Michigan State University, East Lansing, MI 48824, USA
[4] Tsung-Dao Lee Institute, Shanghai Jiao Tong University, No. 1 Lisuo Road, Pudong New Area, Shanghai, People's Republic of China

## Abstract

We present comprehensive photometric and spectroscopic observations of SN 2022eyw, a luminous member of the Type Iax supernova (SN Iax) subclass. SN 2022eyw reached a peak absolute magnitude of $M_g = -17.80 \pm 0.15$ mag and exhibited a rise time of ∼15 days, placing it among the brighter Iax events. The bolometric light curve indicates a synthesized $^{56}$Ni mass of $0.11 \pm 0.01\, M_\odot$, with an estimated ejecta mass of $0.79 \pm 0.09\, M_\odot$ and kinetic energy of $0.19 \times 10^{51}$ erg. The spectral evolution from −8 to +110 days past maximum reveals features characteristic of bright SNe Iax, including a transition from Fe III to Fe II dominance, moderate expansion velocities, and a lack of strong C II absorption. TARDIS spectral modeling of the early-phase spectra indicates well-mixed ejecta dominated by Fe-group elements. In addition, traces of unburnt carbon are detected, pointing to incomplete burning as expected in pure deflagration models. The late-time spectral evolution shows a blend of permitted and forbidden lines. Comparison with deflagration models suggests that SN 2022eyw originated from a partial deflagration of a Chandrasekhar-mass white dwarf, with explosion properties intermediate between the N3-def and N5-def models. These observations support pure deflagration of a CO white dwarf as a viable explosion mechanism for its luminous members.

## 1. Introduction

Type Ia supernovae (SNe Ia) are thermonuclear explosions of carbon–oxygen (CO) white dwarfs in binary systems, occurring when the white dwarf gains mass from a companion and grows toward the Chandrasekhar limit, leading to runaway nuclear burning and the complete disruption of the star (D. Maoz et al. 2014; S. W. Jha et al. 2019). Their luminosity and spectral homogeneity allow them to be described by a one-parameter family (M. M. Phillips 1993; M. M. Phillips et al. 1999) and make them valuable cosmological distance indicators. Type Iax supernovae (SNe Iax; W. Li et al. 2003; R. J. Foley et al. 2013; S. W. Jha 2017) constitute a low-energy subclass of thermonuclear explosions, widely interpreted as the incomplete disruption of the progenitor white dwarf, potentially leaving behind a bound remnant (G. C. Jordan et al. 2012; M. Kromer et al. 2013). Constituting about 15%–30% of SNe Ia (R. J. Foley et al. 2013; S. Srivastav et al. 2022), SNe Iax display a wide diversity of observational characteristics compared to the (nearly) uniform properties of SNe Ia (C. M. McClelland et al. 2010; M. R. Magee et al. 2016; M. Singh et al. 2023). The luminosity of SNe Iax varies widely, with absolute magnitudes ranging from $M_g = -11.66$ mag for SN 2021fcg (V. R. Karambelkar et al. 2021) to $M_g = -18.4$ mag for SN 2012Z (M. D. Stritzinger et al. 2015). Insufficient observations around maximum in most cases hinder an accurate determination of the rise time to maximum in SNe Iax. However, for cases in which sufficient data is available during the premaximum phase and maximum, the light curves are found to rise faster than normal SNe Ia (M. R. Magee et al. 2016, 2017; S. W. Jha 2017; L. Li et al. 2018).

SNe Iax are also distinguished by the nature of their host galaxies. They are predominantly hosted by late-type, star-forming, low-mass to intermediate-mass galaxies, and are rarely found in early-type or passive galaxies (R. J. Foley et al. 2013; J. D. Lyman et al. 2013; S. W. Jha 2017). Environmental analyses and integral-field spectroscopy further show that SNe Iax trace regions of relatively recent star formation, suggesting comparatively short delay times, a result that is also supported by stellar-population age studies of their host galaxies (J. D. Lyman et al. 2018; T. Takaro et al. 2020).

At early times, SNe Iax show notable departures from the spectra of normal SNe Ia, such as strong Fe III and Fe II features and unusually weak Si II absorption (M. M. Phillips et al. 2007; R. J. Foley et al. 2013). These characteristics resemble those of the SN 1991T–like subclass of SNe Ia, a group of luminous Ia events distinguished by prominent Fe III lines and suppressed Si II and Ca II features in their premaximum spectra (A. V. Filippenko et al. 1992; M. M. Phillips et al. 1992). While SNe Iax share this spectral morphology, they exhibit much lower expansion velocities—typically ∼2000–8000 km s$^{-1}$ compared to the significantly higher velocities in 91T-like and normal SNe Ia (S. W. Jha 2017). The C II absorption associated with unburned carbon in the outer layers is generally faint or absent in the more luminous SNe Iax (D. K. Sahu et al. 2008; M. D. Stritzinger et al. 2015; L. Tomasella et al. 2016; A. Dutta et al. 2022; M. Singh et al. 2024), whereas it is clearly detected in the lower-luminosity members of the class (R. J. Foley et al. 2009; M. D. Stritzinger et al. 2014; S. Srivastav et al. 2020, 2022;

L. Tomasella et al. 2020; M. Singh et al. 2023). At late times, SNe Iax develop a mixture of permitted and forbidden emission lines (R. J. Foley et al. 2016).

Given their lower energy output, SNe Iax were thought to result from a different explosion mechanism than the one responsible for typical SNe Ia. Several explosion scenarios have been proposed for SNe Iax, such as deflagration of a CO white dwarf (M. Fink et al. 2014) or carbon–oxygen–neon (CONe) white dwarf (X. Meng & P. Podsiadlowski 2014; M. Kromer et al. 2015), CO and oxygen–neon (ONe) white dwarf mergers (R. Kashyap et al. 2018), mergers of a neutron star/black hole and a white dwarf (A. Bobrick et al. 2022), and many others. Of the various explosion scenarios, deflagration of a CO white dwarf most successfully reproduces the observed characteristics of SNe Iax, particularly those at the luminous end ($M_r \leqslant -17.1$ mag; M. Singh et al. 2023) of the population (M. D. Stritzinger et al. 2015; A. Dutta et al. 2022; M. Singh et al. 2022, 2024). Deflagration of a CONe white dwarf appears to be partially consistent with the characteristics of faint ($M_r \geqslant -14.64$ mag; M. Singh et al. 2023) SNe Iax.

Pre-explosion images provide critical insights into the progenitor systems. Based on archival Hubble Space Telescope images obtained prior to explosion, C. McCully et al. (2014) suggested that SN 2012Z originated from a progenitor system consisting of a CO white dwarf and a helium star companion. Subsequent observations (C. McCully et al. 2022; M. Schwab et al. 2025) further support the survival of the helium star companion, along with the presence of a bound remnant. R. J. Foley et al. (2014) detected red excess at later phases from the location of SN 2008ha, potentially arising from either a bound remnant or the companion star. A progenitor system with a CO white dwarf and a helium star progenitor has also been suggested for the SNe Iax SN 2014dt (R. J. Foley et al. 2015) and SN 2020udy (K. Maguire et al. 2023). The possibility of finding a bound remnant in the case of SNe Iax makes it crucial to identify and perform detailed studies of more such events. This will help in further understanding white dwarf explosions. Besides SNe Ia, the explosive nucleosynthetic yields from SNe Iax are important for galactic chemical evolution models (C. Kobayashi et al. 2020) and hence more explosion models and detailed radiative transfer studies are required for these events.

This paper presents an extensive analysis of the Type Iax SN 2022eyw. SN 2022eyw was discovered by the Pan-STARRS group (K. C. Chambers et al. 2022) using the Pan-STARRS2 telescope on 2022 March 22 at 11:04:36 UT (JD = 2459660.96) at a magnitude of 19.66 in the $i - P1$ filter. It was classified as an SN Iax by C. Balcon (2022) based on a spectrum obtained on 2022 March 25 (JD = 2459664.48). The host galaxy of the supernova (SN) is MCG +11-16-003 at a redshift of $z \approx 0.0099$. It is classified as an Sdm C-type galaxy (H. B. Ann et al. 2015), a system that exhibits very late-type, star-forming characteristics, consistent with the environments commonly associated with SNe Iax. Table 1 lists the essential observational parameters of SN 2022eyw and its host galaxy.

The photometric and spectroscopic observations of SN 2022eyw, along with the data reduction procedures, are described in Section 2. In Section 3, we discuss the adopted distance, extinction, and explosion epoch of the SN. Section 4 presents a detailed analysis of the light curves, and a comparison with synthetic light curves generated from pure deflagration models. The spectroscopic evolution is discussed in Section 5, including spectral modeling using the radiative transfer code TARDIS to interpret the physical conditions and chemical composition of the ejecta. Finally, the main results and conclusions of the study are summarized in Section 6.

**Table 1**
Parameters of SN 2022eyw and Its Host Galaxy

| SN 2022eyw | |
|---|---|
| R.A. (J2000) | $\alpha = 12^{h}43^{m}59\overset{s}{.}970$ [a] |
| Decl. (J2000) | $\delta = +62^{\circ}19'48\overset{''}{.}29$ [a] |
| Discovery date | 2022 Mar 22 11:04:36 UT (JD = 2459660.96153) [a] |
| Last nondetection | 2022 Mar 21 12:28:28 UT (JD = 2459660.01977) [b] |
| Explosion epoch | 2022 Mar 20 21:21:36 UT (JD = $2459659.39^{+0.24}_{-0.27}$) [c] |
| $B$-band maximum | 2022 Apr 3 01:12:00 UT (JD = 2459672.55 ± 0.13) [c] |
| $\Delta m_{15}(B)$ | 1.46 ± 0.05 mag [c] |
| Galaxy reddening | $E(B - V) = 0.012 \pm 0.0003$ mag [d] |
| Host reddening | $E(B - V) = 0.056 \pm 0.005$ mag [c] |
| $^{56}$Ni mass | $M_{Ni} = 0.11\ M_{\odot}$ [c] |
| Ejected mass | $M_{ej} = 0.79\ M_{\odot}$ [c] |
| Kinetic energy | $E_k = 1.97 \times 10^{51}$ erg [c] |
| **Host Galaxy: MCG +11-16-003** | |
| Morphological type | Sdm C [e] |
| R.A. (J2000) | $\alpha = 12^{h}43^{m}59\overset{s}{.}954$ [f] |
| Decl. (J2000) | $\delta = +62^{\circ}19'59\overset{''}{.}71$ [f] |
| Redshift | $z = 0.0099 \pm 0.00006$ [c] |
| Distance modulus | $\mu = 33.04 \pm 0.15$ mag [c] |

**Notes.**
[a] K. C. Chambers et al. (2022).
[b] https://www.wis-tns.org/object/2022eyw.
[c] This work.
[d] E. F. Schlafly & D. P. Finkbeiner (2011).
[e] H. B. Ann et al. (2015).
[f] K. N. Abazajian et al. (2009).

## 2. Observations and Data Reduction

### 2.1. Photometry

Observations of SN 2022eyw were carried out using the 2 m Himalayan Chandra Telescope (HCT; T. P. Prabhu & G. C. Anupama 2010) at the Indian Astronomical Observatory, Hanle. The telescope is equipped with the Himalayan Faint Object Spectrograph Camera (HFOSC), which enables both optical imaging and low-resolution to medium-resolution spectroscopy. Data collection commenced on 2022 March 25, and continued until 2022 June 1. For photometric calibration, observations of Landolt standard fields PG0918 +029, PG0942-029, PG1047+003, and PG1528+062 were performed on 2022 March 26. Bright stars in the SN field were then calibrated so as to use them as secondary standards for calibrating the SN magnitude (Figure 1).

Photometric reduction involved correcting the raw CCD frames for bias and flat-field effects and aligning the frames with respect to a reference frame. As the SN was embedded inside the host galaxy, to remove the effect of contamination in the SN brightness due to varying galaxy background, subtraction of the galaxy template image was essential. The template images were observed with the same instrumental setup more than 2 yr after the discovery, when the SN faded below the detection limit. Template subtraction was performed following the standard procedure, wherein the point-spread functions (PSFs) of the science and reference images were matched, and the background levels were scaled appropriately. The reference image was then subtracted from the science frame to generate the difference image. Aperture photometry on these difference images was subsequently carried out to estimate the SN magnitudes. Since the host galaxy was very faint in the *U* band and was close to the sky background in the image, template subtraction was not done for this band. Finally, the nightly zero-points were derived using secondary standard stars in the SN field, and the SN magnitude was brought to the standard system. The associated error in the SN magnitude was estimated by taking into account the fit error provided by IRAF (D. Tody 1986) and uncertainty in nightly zero-points.

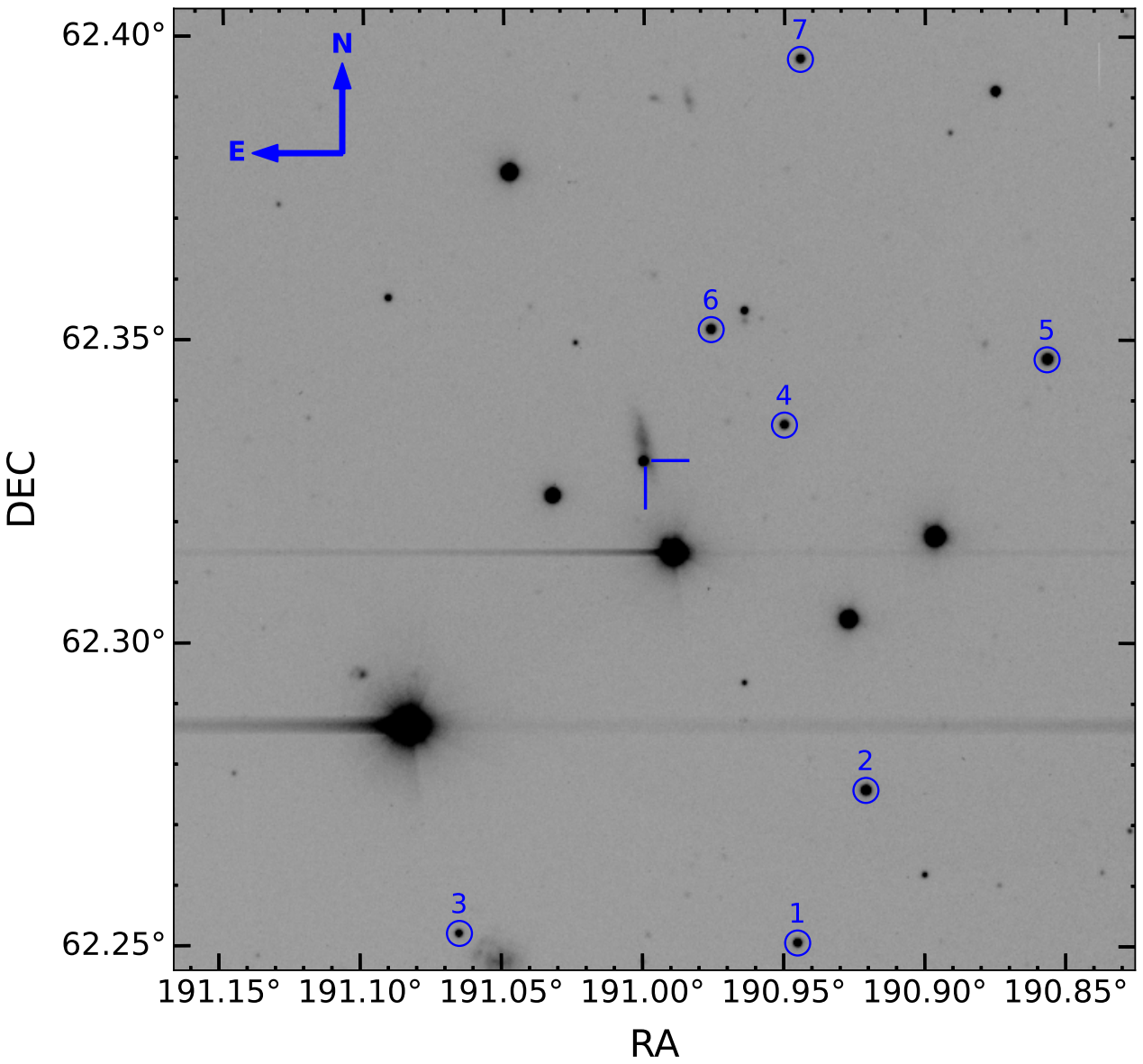

**Figure 1.** *B*-band image of the SN 2022eyw field, with a field of view of approximately $9'.5 \times 9'.5$. The image was taken using the 2 m HCT on 2022 March 26 with an exposure time of 360 s. The SN is marked with blue crosshairs, and the secondary standard stars used for photometric calibration are enclosed with blue circles.

Additional photometric observations were conducted in the Sloan Digital Sky Survey (SDSS) $g'$, $r'$, $i'$, and $z'$ filters using the 0.7 m GROWTH-India Telescope (GIT), a fully robotic telescope located at the Indian Astronomical Observatory, Hanle (H. Kumar et al. 2022a). GIT is optimized for rapid follow-up of transient events and is equipped with a $4096 \times 4108$ pixel Andor XL 230 camera. Observations of SN 2022eyw with GIT commenced on 2022 March 25, and continued until 2022 July 11. The telescope was operated in targeted mode to maximize the observation cadence. For photometric calibration, the Pan-STARRS catalog (H. A. Flewelling et al. 2020) was used to determine zero-points. Pan-STARRS reference images were used for host galaxy subtraction. Image subtraction was performed using PYZOGY, which is a Python implementation of the ZOGY algorithm (B. Zackay et al. 2016), ensuring precise background subtraction and transient flux recovery. The PSF model generated by PSFex (E. Bertin 2011) was employed to perform photometry on the difference images, allowing accurate extraction of the SN's magnitudes. All of these steps are integrated into the GIT Image Subtraction Pipeline (H. Kumar et al. 2022b), which was used to obtain the final apparent magnitudes in the $g'$, $r'$, $i'$, and $z'$ bands.

SN 2022eyw was also monitored by the Zwicky Transient Facility (ZTF; E. C. Bellm et al. 2019) in the *g* and *r* bands. The ZTF observations span the period of 2022 April 1 to 2022 June 30, providing additional coverage of the SN's photometric evolution. The photometric data from ZTF were obtained from the public archive and incorporated into the analysis to supplement light-curve construction and improve the temporal coverage.

SN 2022eyw was observed with Swift/UVOT (P. W. A. Roming et al. 2005) on board the Neil Gehrels Swift Observatory (N. Gehrels et al. 2004), from 2022 March 7 in the *UVW*2, *UVM*2, *UVW*1, *U*, *B*, and *V* filters. We downloaded the publicly accessible data for SN 2022eyw from the Swift portal,[5] which was available under two different target IDs (00015099 and 03111677). The data reduction and photometry were performed using the standard procedure, utilizing packages included in High Energy Astrophysics Software (HEASOFT, v6.27) along with the latest calibration database for the UVOT instrument, following the prescription from T. S. Poole et al. (2008) and P. J. Brown et al. (2009).

### 2.2. Spectroscopy

Spectroscopic observations were conducted from 2022 March 26 to 2022 July 23, spanning a total of 119 days. Data were acquired using HFOSC with grisms Gr7 (3500–7800 Å) and Gr8 (5200–9100 Å). Data reduction was performed using standard tasks available within IRAF. The spectral images were bias-subtracted and flat-fielded, and one-dimensional spectra were extracted using the optimal extraction method available within IRAF (K. Horne 1986). Wavelength calibration was carried out to convert the pixel scale to wavelength scale using a dispersion solution obtained with the help of FeAr and FeNe arc lamp spectra. Since the spectra were extracted in multispec format, the sky spectra were also obtained and were used to verify the wavelength calibration by checking sky lines and if necessary, minor wavelength shifts were applied. Spectrophotometric standard stars, observed on the same nights as the SN, were used for flux calibration to bring them to a relative flux scale. The flux-calibrated spectra in the two regions were then combined using a weighted mean to produce the final spectrum. Finally, the combined spectra were scaled with respect to the photometric flux to bring them to an absolute flux scale. Additionally, the spectra were corrected for a redshift of $z = 0.0099$. Dereddening was applied to account for extinction caused by both the host galaxy and the Milky Way, using a total $E(B-V) = 0.068 \pm 0.005$ mag (refer to Section 3.1 for details).

## 3. Distance, Extinction, and Explosion Epoch

### 3.1. Distance and Extinction

The host galaxy's redshift was determined using the radial velocity (*cz*) corrected for the Local Group's infall toward the Virgo cluster (J. R. Mould et al. 2000).[6] The estimated redshift is $z = 0.0099 \pm 0.00006$, which is consistent with the value

5 https://www.swift.ac.uk/

6 https://ned.ipac.caltech.edu/

obtained using the narrow H$\alpha$ emission line from the host galaxy, seen in the spectra of SN 2022eyw.

The reddening due to the host galaxy was estimated using the equivalent width of the Na ID absorption lines. In the near-maximum spectrum of the SN, taken at $-1.8$ days from the $g'$-band maximum, Na ID lines were detected at the host's redshift with a pseudo equivalent width (pEW) of $0.35 \pm 0.03$ Å. Applying the empirical relation $E(B-V) = 0.16 \times \mathrm{pEW(Na\ ID)}$ (M. Turatto et al. 2003), a host galaxy reddening of $E(B-V)_{\mathrm{host}} = 0.056 \pm 0.005$ mag was estimated. The Galactic reddening along the line of sight is $E(B-V)_{\mathrm{Gal}} = 0.012 \pm 0.0003$ mag (E. F. Schlafly & D. P. Finkbeiner 2011), leading to a total extinction of $0.068 \pm 0.005$ mag. Adopting $R_V = 3.1$, we got $A_V = 0.211 \pm 0.015$. The `extinction` module of Python was then used to calculate the extinction in other bands, according to the Fitzpatrick and Massa extinction model (E. L. Fitzpatrick & D. Massa 2007).

Assuming a cosmology with $H_0 = 73$ km s$^{-1}$ Mpc$^{-1}$, $\Omega_\Lambda = 0.73$, and $\Omega_M = 0.27$, a distance modulus of $\mu = 33.04 \pm 0.15$ mag (and thus a distance of $40.6 \pm 2.8$ Mpc) was derived. Using this distance and correcting for the total extinction, the peak absolute magnitude in the $g'$ band was estimated as $M_g = -17.80 \pm 0.15$ mag.

### 3.2. Explosion Epoch

The explosion epoch of the SN was estimated using two independent methods: (i) by averaging the Julian Dates (JDs) of the last nondetection and discovery, and (ii) by fitting the bolometric light curve with the Arnett model (W. D. Arnett 1982).

The last nondetection of SN 2022eyw was reported on 2022 March 21 ($\mathrm{JD_{last}} = 2459660.02$) with a limiting magnitude of 18.33 in the ATLAS "o" filter by the ATLAS-HKO telescope (as listed in the Transient Name Server[7]) and the discovery JD ($\mathrm{JD_{disk}}$) was reported to be JD 2459660.96 (K. C. Chambers et al. 2022). Assuming a uniform probability distribution for the explosion time within this interval, the explosion epoch was estimated as $\mathrm{JD_{exp}} = 2459660.49 \pm 0.47$, the arithmetic mean of the last nondetection and discovery epochs.

Alternatively, by fitting the bolometric light curve with the Arnett model (see Section 4.3.2), an explosion epoch of $\mathrm{JD} = 2459659.39^{+0.24}_{-0.27}$ was obtained. This method provides a more physically motivated estimate by modeling the radioactive heating and diffusion processes that govern the luminosity evolution of the SN. Both methods yielded consistent results within their respective uncertainties, with a maximum discrepancy of 1.0 day. Given the consistency between these methods, the Arnett model–derived explosion epoch was adopted for subsequent analysis and used to estimate the rise times for different bands, tabulated in Table 2.

## 4. Light Curve and Color Curve

### 4.1. Light-curve Properties and Analysis

The photometric evolution of SN 2022eyw was followed in Bessell's $U, B, V, R$ bands and the SDSS $g', r', i', z'$ bands. While GIT employs the SDSS $u'g'r'i'z'$ filters and ZTF uses its own custom $g$ and $r$ filters, the differences in their respective filter response functions and effective wavelengths are minimal. Therefore, the data from both instruments were combined to construct the $g'$- and $r'$-band light curves. For the $i'$ and $z'$ bands, only GIT observations were used to construct the light curves. The $U$-, $B$-, and $V$-band light curves were generated using HCT data, supplemented by a few additional data points from the Swift/UVOT instrument. The Swift observations in the $UVW2$, $UVM2$, and $UVW1$ filters were not included as they yielded nearly constant magnitudes with large uncertainties, consistent with background noise. In the case of the $R$ band, the light curve has been constructed exclusively from HCT observations.

Figure 2 presents the light curves of the SN in the $U$, $B$, $g'$, $V$, $r'$, $R$, $i'$, and $z'$ bands. The dense photometric coverage around the peak enables reliable determination of the key light-curve parameters. We estimated the peak magnitude, epoch of maximum, and $\Delta m_{15}$ values in each band using spline fits to the observed data. To estimate the uncertainties in these measurements, we conducted a Monte Carlo simulation where random Gaussian noise, centered at zero with a standard deviation equal to the photometric error of each data point, was added to the magnitudes. This procedure was repeated for a sufficiently large number of iterations until the resulting distributions of the fitted parameters converged. The mean of the resulting distribution for each parameter was adopted as its final value, with the corresponding standard deviation as its uncertainty.

The light-curve evolution of SN 2022eyw across multiple bands exhibits characteristics typical of SNe Iax. A wavelength-dependent trend is evident in the light-curve evolution of SN 2022eyw, with the bluer bands peaking earlier than the redder ones. The SN reached its $B$-band maximum magnitude of $15.71 \pm 0.02$ mag at JD 2459672.55 and its $g'$-band maximum of $15.50 \pm 0.02$ mag at JD 2459674.00. The $U$ band peaked 1.14 days before the $B$-band maximum, while the $V$ and $R$ bands peaked $+2.5$ and $+5.12$ days later, respectively. A similar progression is observed in the SDSS bands, where the $r'$, $i'$, and $z'$ light curves peaked at $+4.14$, $+5.85$, and $+7.24$ days after the $g'$-band maximum. This systematic shift in peak epochs is consistent with the expected thermal evolution of the expanding SN ejecta. Similar behavior has been reported in other SNe Iax, such as SN 2005hk (M. M. Phillips et al. 2007; D. K. Sahu et al. 2008) and SN 2011ay (R. J. Foley et al. 2013).

The peak absolute magnitudes of SN 2022eyw span $-18.20(U)$ to $-17.58(z')$ mag. In the $V$ band, the SN attains a peak absolute magnitude of $-17.80 \pm 0.15$ mag, which is comparable to SN 2005hk ($M_V \approx -18.08$; M. M. Phillips et al. 2007) and SN 2012Z ($M_V \approx -18.50$; M. D. Stritzinger et al. 2015), but brighter than low-luminosity events like SN 2008ha ($M_V \approx -14.21$; R. J. Foley et al. 2009) or SN 2010ae ($M_V \approx -14.11$; M. D. Stritzinger et al. 2014), thus placing it among the brighter members of the Iax subclass. The rise time, which increases systematically from blue to red bands—ranging from ∼12.2 days in $U$ to ∼22.0 days in $z'$—suggests a temperature evolution that aligns with expectations from radiative diffusion models, where the optical peak shifts to longer wavelengths as the ejecta expands and cools.

The postmaximum decline rate, quantified by the $\Delta m_{15}$ parameter, shows a clear trend of faster decline in the bluer bands ($\Delta m_{15}(U) = 1.45$) and slower decline in the redder bands ($\Delta m_{15}(z') = 0.44$) (refer to Table 2). The decline rate is similar to other well-studied SNe Iax like

[7] https://www.wis-tns.org/object/2022eyw

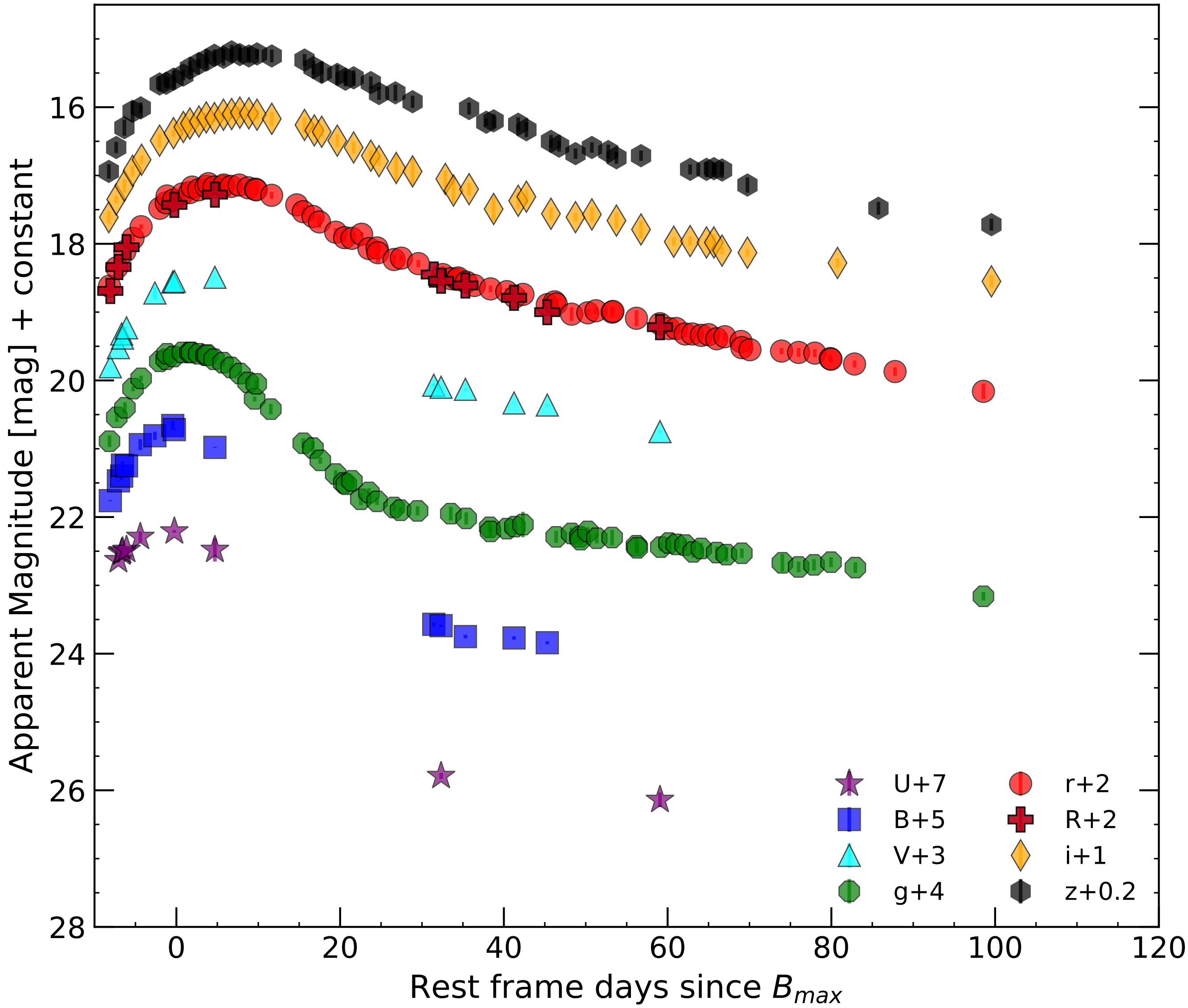

**Figure 2.** Light-curve evolution of SN 2022eyw in $UBg'Vr'i'z'$ bands. The $B$-band maximum is used to calculate the phase which is plotted on the $x$-axis. The $y$-axis shows the apparent magnitudes in the respective bands, shifted by a constant for a clear representation. The light curve in each band is plotted with its corresponding error bars. However, the uncertainties are relatively small, causing the error bars to be mostly concealed behind the data markers. All the magnitudes are in the Vega system. We have used the Vega–AB magnitude conversion factors (M. R. Blanton & S. Roweis 2007) to convert the $g'r'i'z'$ magnitudes to the Vega system from the AB system.

**Table 2**
Photometric Parameters of SN 2022eyw

| Filter | $\lambda_{\rm eff}$ (Å) | JD (Max) | $m_\lambda^{\rm max}$ | $\Delta m_{15}(\lambda)$ | $t_R$ (days) | $M_\lambda^{\rm max}$ |
|---|---|---|---|---|---|---|
| $U$ | 3663.6 | 2459671.41 ± 0.82 | 15.18 ± 0.04 | 1.45 ± 0.28 | 12.12 | −18.18 ± 0.16 |
| $B$ | 4363.2 | 2459672.55 ± 0.13 | 15.71 ± 0.02 | 1.46 ± 0.05 | 13.32 | −17.61 ± 0.15 |
| $g'$ | 4734.2 | 2459674.00 ± 0.33 | 15.50 ± 0.02 | 1.43 ± 0.06 | 14.77 | −17.80 ± 0.15 |
| $V$ | 5445.8 | 2459675.05 ± 0.49 | 15.47 ± 0.02 | 0.85 ± 0.05 | 15.82 | −17.79 ± 0.15 |
| $r'$ | 6238.4 | 2459678.14 ± 0.18 | 15.32 ± 0.01 | 0.62 ± 0.01 | 18.91 | −17.90 ± 0.15 |
| $R$ | 6414.2 | 2459677.67 ± 0.49 | 15.25 ± 0.01 | 0.64 ± 0.01 | 18.44 | −17.96 ± 0.15 |
| $i'$ | 7751.1 | 2459679.85 ± 0.34 | 15.47 ± 0.02 | 0.51 ± 0.02 | 20.62 | −17.69 ± 0.15 |
| $z'$ | 9106.7 | 2459681.24 ± 0.38 | 15.56 ± 0.03 | 0.44 ± 0.02 | 22.01 | −17.57 ± 0.15 |

SN 2005hk ($\Delta m_{15}(B) \approx 1.68$; D. K. Sahu et al. 2008), SN 2012Z ($\Delta m_{15}(B) \approx 1.43$; M. D. Stritzinger et al. 2015), SN 2020rea ($\Delta m_{15}(B) \approx 1.61$; M. Singh et al. 2022), and SN 2020udy ($\Delta m_{15}(B) \approx 1.36$; M. Singh et al. 2024).

Notably, SN 2022eyw does not exhibit a secondary maximum in the redder bands (e.g., $R$, $i'$, or $z'$), a feature commonly seen in normal SNe Ia. This secondary peak is typically caused by the recombination of iron-group elements (IGEs) and associated opacity changes in the ejecta. However, in SNe Iax, the lower ejecta mass, mixed abundance structure, and generally lower ionization state tend to suppress or smooth out this feature (D. Kasen 2006; F. Lach et al. 2022).

The early-time light curve is powered by the radioactive decay of $^{56}$Ni → $^{56}$Co, while the postpeak decline is governed by the $^{56}$Co → $^{56}$Fe decay chain. The observed slower decline in redder bands as compared to bluer bands reflects the cooling and recombination of the ejecta over time. As the ejecta cools and Fe III recombines to Fe II, numerous Fe II/Co II lines blanket the $B$ band and reemit thermalized energy from radioactive decay at longer wavelengths (D. Kasen & S. E. Woosley 2007). The observed trend between rise time, decline rate, and peak magnitude is consistent with a broader correlation seen in SNe Iax, where brighter events tend to have longer rise times and shallower postmaximum declines, reflecting a higher $^{56}$Ni mass and more extended diffusion timescales (R. J. Foley et al. 2013; M. R. Magee et al. 2016).

### 4.2. Color-curve Evolution

The color evolution of SN 2022eyw is shown in Figure 3. The reddening-corrected color curves of SN 2022eyw are plotted with the color evolution of some other well-studied SNe Iax. The evolution of $(B-V)$, $(g'-r')$, $(V-R)$, and $(r'-i')$ follows a smooth, monotonic reddening trend, due to expansion and cooling of the ejecta, similar to other Type Iax events. These consistent color trends suggest a broadly homogeneous thermal evolution across the subclass, even though the timescales and exact color values may vary. Interestingly, the colors are also redder in the premaximum phase, which can be attributed to longer photon diffusion times for bluer wavelengths compared to redder ones. As the SN reaches maximum light, when trapped radiation escapes more freely, the colors appear bluer, reflecting the peak of thermal emission.

### 4.3. Bolometric Light-curve Analysis

#### 4.3.1. Comparison and Analysis of Pseudobolometric Light Curve

The bolometric light curve of SN 2022eyw was constructed using the Python-based tool `SuperBol` (M. Nicholl 2018), incorporating photometry in the $U$, $B$, $V$, $R$, $g'$, $r'$, $i'$, and $z'$ bands. The extinction and distance modulus, as discussed in Section 3.1, were provided as inputs. `SuperBol` converts the dereddened magnitudes into monochromatic fluxes using standard zero-points and effective wavelengths. These fluxes are used to construct spectral energy distributions at each observational epoch.

To compute the pseudobolometric luminosity, `SuperBol` integrates the flux across the observed wavelength range using the trapezoidal rule. To account for incomplete coverage, `SuperBol` interpolates missing photometric points using polynomial fits to the observed light curves in each band. For epochs where certain filters are not observed, it performs color-based extrapolation using the closest available colors from adjacent epochs, assuming they remain roughly constant. The code also propagates uncertainties based on the flux errors and filter bandwidths to derive the luminosity error at each epoch.

For comparison with other SNe Iax, whose bolometric light curves are often reported using only the $BgVri$ filter set, we constructed a $BgVri$-based pseudobolometric curve for SN 2022eyw. As shown in Figure 4, the overall shape of the bolometric light curve of SN 2022eyw is consistent with that of bright Type Iax events such as SN 2005hk, SN 2012Z, and SN 2020udy. The inset in Figure 4 shows the pseudobolometric light curve after including the $U$ and $z'$ bands, which apparently increases the pseudobolometric luminosity. The peak luminosities were estimated to be $L_{\rm peak}^{BgVri} = (1.86 \pm 0.06) \times 10^{42}$ erg s$^{-1}$ and $L_{\rm peak}^{UBgVrRiz} = (2.52 \pm 0.06) \times 10^{42}$ erg s$^{-1}$, respectively. Thus at peak, the $U$ and $z'$ bands together contribute ∼26% of the total pseudobolometric luminosity, underscoring the significance of wavelength coverage when constructing bolometric light curves.

It is important to note that the pseudobolometric light curve constructed from the observed bands does not include flux in the unobserved UV and IR regions. For SNe Iax, a well-defined correction factor for these missing regions does not exist due to the intrinsic diversity and limited wavelength coverage in most events. However, several well-observed objects provide a useful empirical estimate of UV and IR contribution near the peak phase of the light curve. At peak, the combined UV and IR contribution is often assumed to be around 35% (L. Tomasella et al. 2016; M. Singh et al. 2025). This estimate is fairly consistent with the findings of A. Dutta et al. (2022) and S. Srivastav et al. (2020), who compared pseudobolometric and blackbody-corrected luminosities and found that the optical contribution accounts for approximately 62% and 69% of the total peak luminosity for SN 2020sck and SN 2019gsc, respectively.

#### 4.3.2. Analytical Estimate of Radioactive $^{56}$Ni and Other Explosion Parameters

The amount of radioactive $^{56}$Ni produced in the explosion was estimated from the $UBgVriz$ pseudobolometric light curve shown in Figure 4. The rate of energy released from the radioactive decay of $^{56}$Ni → $^{56}$Co → $^{56}$Fe is given by D. K. Nadyozhin (1994) as

$$L_R(t) = (6.45 \times 10^{43} e^{-\frac{t}{8.8}} + 1.45 \times 10^{43} e^{-\frac{t}{111.3}}) \times \frac{M_{\rm Ni}}{M_\odot}\ {\rm erg\ s^{-1}}. \quad (1)$$

This equation does not take into account the energy from neutrinos emitted by $^{56}$Ni and $^{56}$Co decay. The luminosity released in the radioactive decay chain ($L_R(t)$) is not entirely deposited in the ejecta. Some of the energy is carried away by the $\gamma$-rays and positrons and hence lost. The output luminosity of an SN is obtained by solving the equation for energy conservation in a diffusive medium, using the luminosity given by Equation (1) as input. The ejecta is assumed to be spherically symmetric, expanding homologously and the radiation energy to be dominant over gas energy. The expression for the output luminosity is given as (see W. D. Arnett 1982; Equation (A1) of S. Valenti et al. 2008; Equation (2) of E. Chatzopoulos et al. 2009)

$$L(t) = M_{\rm Ni} e^{-x^2}\Big[(\epsilon_{\rm Ni} - \epsilon_{\rm Co}) \int_0^x 2z e^{z^2-2zy}\,dz + \epsilon_{\rm Co} \int_0^x 2z e^{z^2-2yz+2zs}\,dz\Big]\Big(1 - e^{-(\frac{t_\gamma}{t})^2}\Big). \quad (2)$$

The factor $\left(1 - e^{-\left(\frac{t_\gamma}{t}\right)^2}\right)$ accounts for the trapping of $\gamma$-rays in the ejecta. Here, $x \equiv t/t_{\rm lc}$, where $t$ is the time since explosion (days) and $t_{\rm lc}$ is the light-curve timescale (days); $y \equiv t_{\rm lc}/(2t_{\rm Ni})$ with $t_{\rm Ni} = 8.8$ days; and $s \equiv [t_{\rm lc}(t_{\rm Co} - t_{\rm Ni})/(2t_{\rm Co}t_{\rm Ni})]$ with $t_{\rm Co} = 111.3$ days. $M_{\rm Ni}$ is the initial Ni mass and $t_\gamma$ is the $\gamma$-ray timescale (days); $\epsilon_{\rm Ni} = 3.9 \times 10^{10}$ erg s$^{-1}$ g$^{-1}$ and $\epsilon_{\rm Co} = 6.8 \times 10^{9}$ erg s$^{-1}$ g$^{-1}$ are the energy generation rates due to the decay of Ni and Co, respectively. The output luminosity in Equation (2) was fit to the integrated *UBgVrRiz* light curve with $t_{\rm exp}$, $t_{\rm lc}$, $t_\gamma$, and $M_{\rm Ni}$ being the fit parameters. We fitted the data up to JD 2459715 ($\sim$40 days since the *B*-band maximum). Beyond this JD a good fit to the peak and the tail part of the light curve simultaneously was not achieved. This is because the model works under the assumption of diffusion approximation, which breaks down in the later epochs. The posterior distribution of the parameters was sampled and the 16th, 50th, and 84th percentiles are reported (refer to Figure 5; see also A. Dutta et al. 2022). The fit to the data gave $JD_{\rm exp} = 245\,9659.39^{+0.24}_{-0.27}$, $M_{\rm Ni} = 0.11^{+0.01}_{-0.01}$ $M_\odot$, $t_{\rm lc} = 12.59^{+0.66}_{-0.51}$ days, and $t_\gamma = 40.41^{+1.09}_{-1.26}$ days.

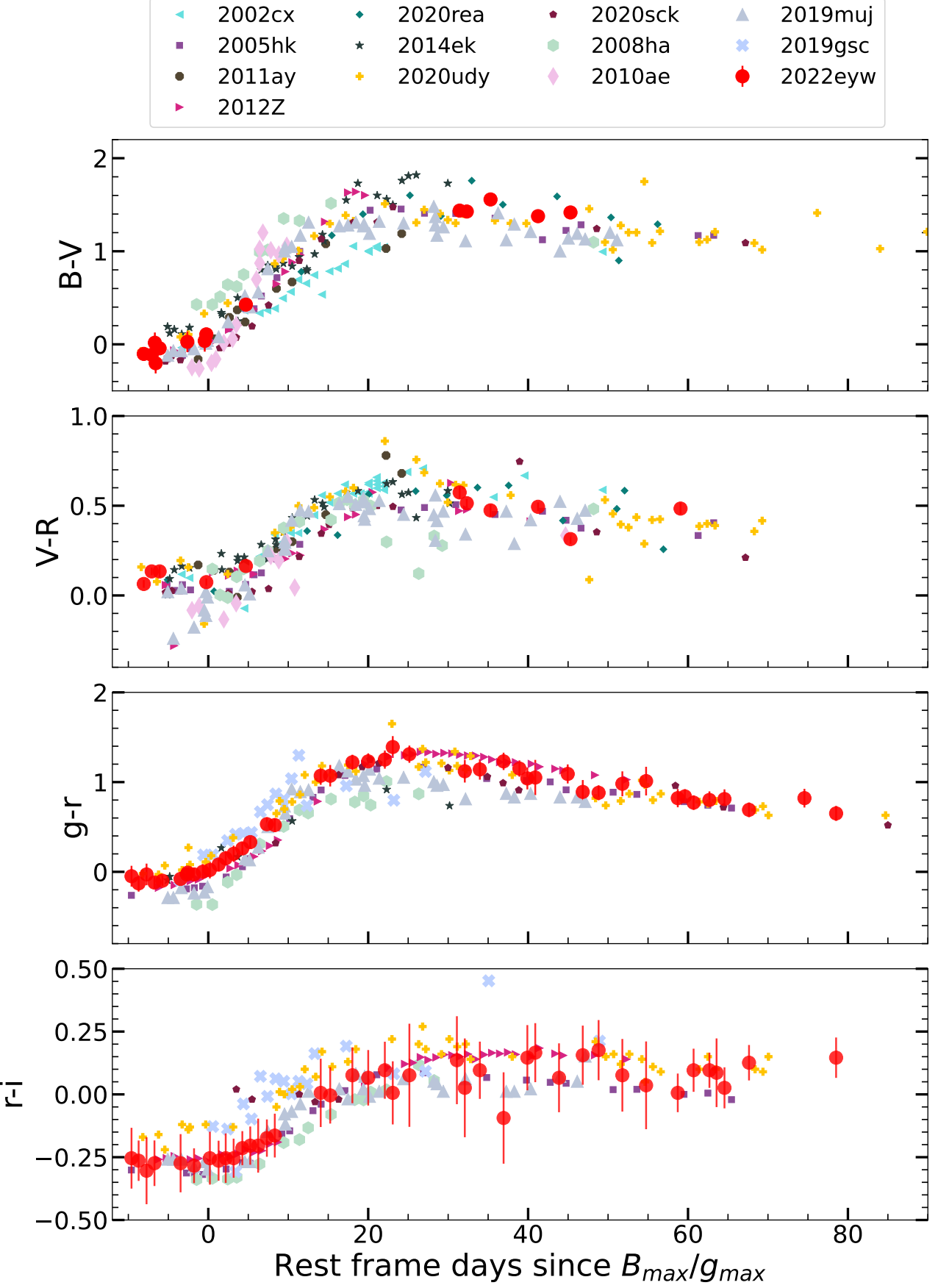

**Figure 3.** $(B - V)$, $(V - R)$, $(g - r)$, and $(r - i)$ color evolution of SN 2022eyw, plotted alongside a comparison sample of other well-observed SNe Iax. All colors are corrected for extinction. Dark-colored small markers are used to represent bright SNe Iax while soft-colored large markers are used to represent faint SNe Iax.

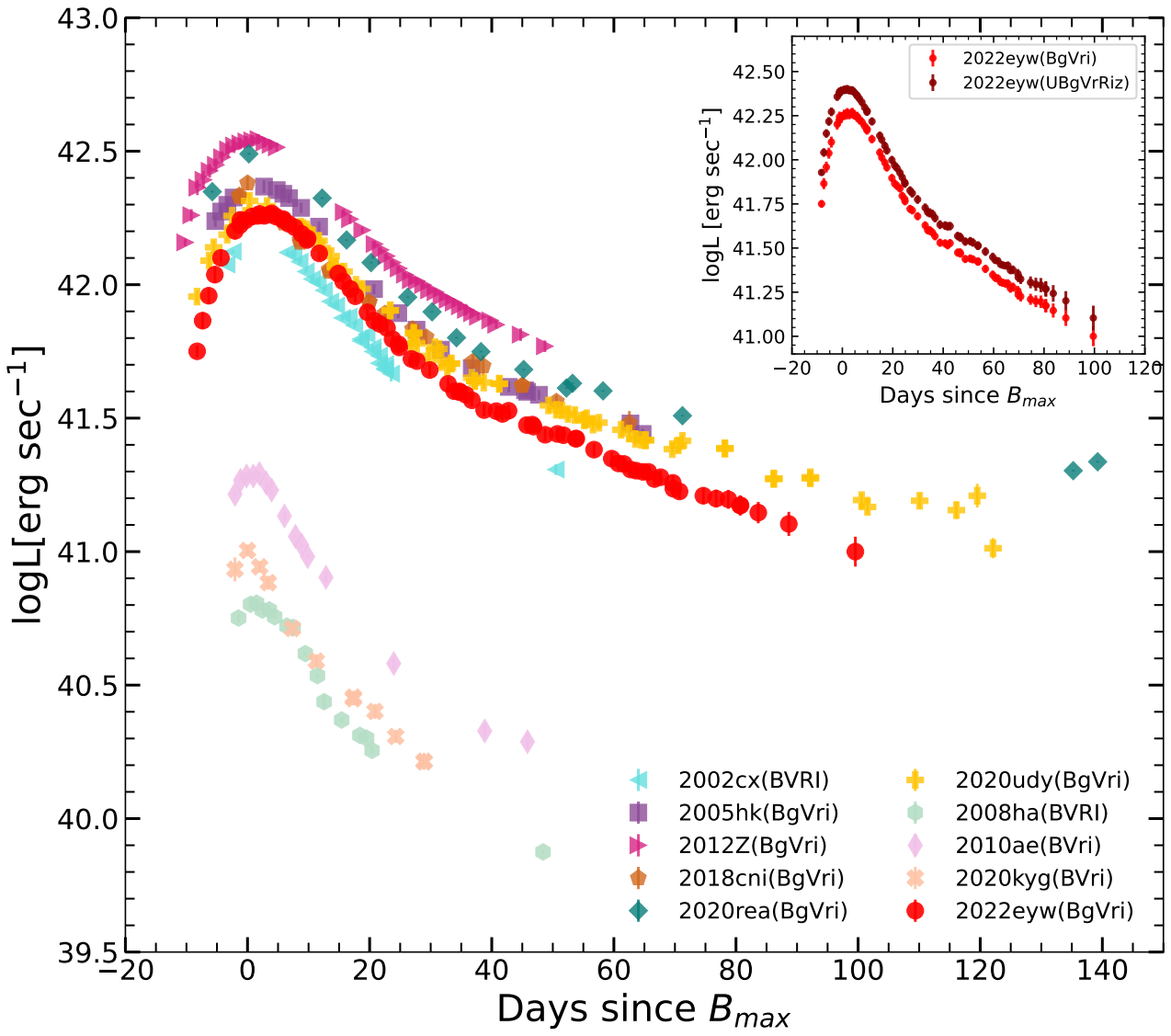

**Figure 4.** Pseudobolometric light curve of SN 2022eyw (red circles) constructed using the *BgVri* filter set, compared with other bright SNe Iax—SN 2002cx, SN 2005hk, SN 2012Z, SN 2018cni, SN 2020rea, and SN 2020udy—and faint ones: SN 2008ha, SN 2010ae, and SN 2020kyg. The inset highlights the difference in luminosity when the $U$ and $z'$ bands are included in the integration, resulting in a higher peak pseudobolometric flux.

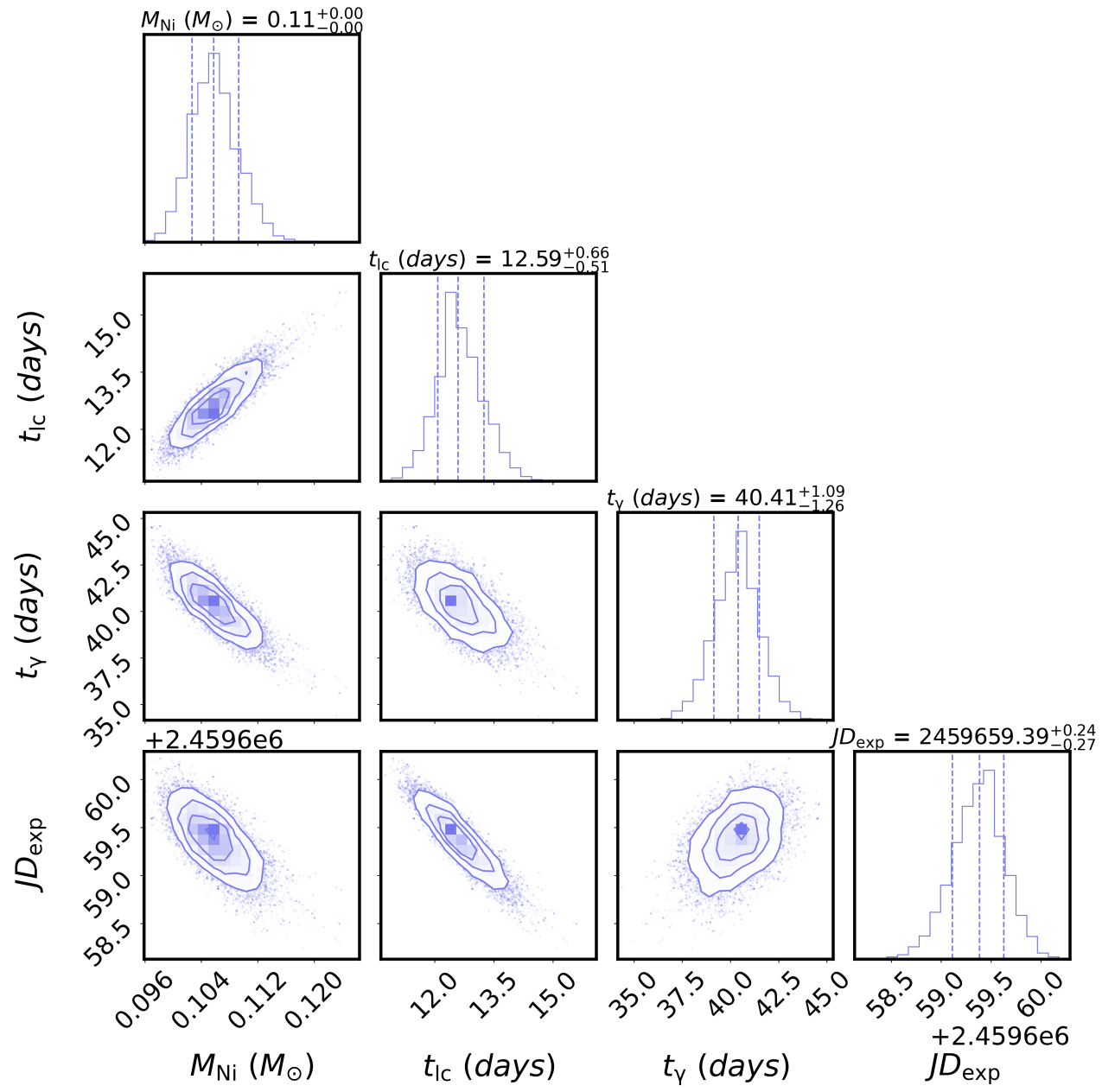

**Figure 5.** One-dimensional (along the diagonal) and two-dimensional projections of the posterior distribution of the parameters of the fit. The 16th, 50th, and 84th percentiles are shown as dashed lines.

As discussed in Section 4.3.1, the combined contribution from unobserved UV and IR flux is commonly estimated to be $\sim$35% near peak brightness. Since our pseudobolometric light curve constructed from the $U$–$z$ bands already includes a substantial portion of this missing flux (about 26%), the remaining $\sim$9% has only a small effect on the derived value of $^{56}$Ni mass and lies well within its uncertainty. For this reason, we adopted the $^{56}$Ni mass obtained from the Markov Chain Monte Carlo fit of Arnett's semianalytical model to the *UBgVrRiz* pseudobolometric light curve as our final estimate.

The ejecta mass ($M_{\rm ej}$) and kinetic energy ($E_{\rm k}$) of the explosion can be obtained by using the relations

$$M_{\rm ej} = 0.5\frac{\beta c}{\kappa} v_{\rm exp} t_{\rm lc}^2 \tag{3}$$

$$E_{\rm k} = 0.3 M_{\rm ej} v_{\rm exp}^2 \tag{4}$$

where $\beta = 13.8$ is a dimensionless constant of integration, $c$ is the speed of light, and $v_{\rm exp}$ and $\kappa$ are the characteristic expansion velocity and opacity of the ejecta, respectively. Assuming a constant optical opacity $\kappa_{\rm opt} = 0.1\ {\rm cm^2\ g^{-1}}$ appropriate for Fe-dominated ejecta (P. A. Pinto & R. G. Eastman 2000a; T. Szalai et al. 2015; S. Srivastav et al. 2020), and considering an expansion velocity of $v_{\rm exp} \approx 6400\ {\rm km\ s^{-1}}$ near maximum light (see Section 5.3), the ejecta mass was estimated as $M_{\rm ej} = 0.79^{+0.08}_{-0.07}\ M_\odot$ and the kinetic energy of explosion as $E_{\rm k} = 0.19^{+0.02}_{-0.01} \times 10^{51}$ erg.

#### 4.3.3. Comparison with Deflagration Models

SNe Iax are widely believed to be the deflagration of a Chandrasekhar-mass CO white dwarf (M. R. Magee et al. 2016; F. Lach et al. 2022). The models developed by M. Fink et al. (2014) simulate such explosions by varying the number of ignition spots (*N*), which control the energy release and the amount of synthesized nickel and hence the luminosity and evolution of the light curve. These models—labeled as N3-def, N5-def, N10-def, and N20-def—span a wide range of explosion energies and are often used as comparison templates for SNe Iax to understand their different behaviors.

To compare SN 2022eyw with these theoretical models, we constructed synthetic bolometric light curves for the N3-def, N5-def, N10-def, and N20-def models using their angle-averaged optical spectral time series. These synthetic spectra provide flux densities at various epochs in units of ${\rm erg\ s^{-1}\ cm^{-2}\ \AA^{-1}}$, assuming a source distance of 10 pc. For each epoch, we numerically integrated the model spectra between 3500 and 9500 Å to calculate the total flux. The resulting fluxes were then converted into bolometric luminosities using the standard luminosity–distance relation. While our observed pseudobolometric light curve spans a broader wavelength range (∼3000–11000 Å), the slight mismatch does not significantly affect the comparison since the objective is to examine the relative shapes and temporal evolution of the light curves.

As shown in Figure 6, the peak of the bolometric light curve and inferred $^{56}$Ni mass for SN 2022eyw lies between the N3-def and N5-def models. The postmaximum decline rate is best matched by the N3-def model. This comparison supports the interpretation that SN 2022eyw arose from a low-energy, partial deflagration event, consistent with the broad theoretical framework of Type Iax explosions.

To further test the consistency of SN 2022eyw with deflagration-based explosion models, we compared its multiband light curves with synthetic photometry from the N3-def, N5-def, N10-def, and N20-def models of M. Fink et al. (2014; Figure 7). Synthetic magnitudes were computed by convolving the model spectra with the corresponding filter transmission curves using `sncosmo`. Since the model assumes the SN to be at a distance of 10 pc, the magnitudes that we get are absolute magnitudes. These were then directly compared with the absolute magnitudes of SN 2022eyw in the same bands.

For SN 2022eyw, the peak magnitudes in all bands fall between the N3-def and N5-def models (refer to Figure 7). This is expected as the early-phase light curves of thermonuclear SNe are primarily powered by the radioactive decay of $^{56}$Ni synthesized during the explosion, and the peak luminosity in each band becomes directly dependent on the amount of $^{56}$Ni produced (C. McCully et al. 2022). The comparison between synthetic light curves from the Fink deflagration models and the observed light curves of SN 2022eyw across the $Bg'VRr'i'$ bands reveals systematic discrepancies that grow with increasing wavelength. In the $B$, $g'$, and $V$ bands, light curves from the Fink model rise faster than the observed light curves. The models also slightly overpredict the postmaximum decline rates in these bands. The discrepancies become more pronounced in the redder $R$, $r'$, and $i'$ bands, where the observed light curves are systematically broader and decline more slowly than their synthetic counterparts. For instance, the N3-def model shows a decline rate of $\Delta m_{15}(g') \approx 1.63$, while the observed $\Delta m_{15}(g')$ for SN 2022eyw is 1.43. This difference in decline rate increases in the $r'$ band, for which the N3-def model gives $\Delta m_{15}(r') \approx 1.19$, compared to the observed value of 0.62 (refer to Table 3). The predicted steeper decline rates in the models suggest insufficient radiation trapping in the ejecta, the effects of which become more prominent in the redder bands.

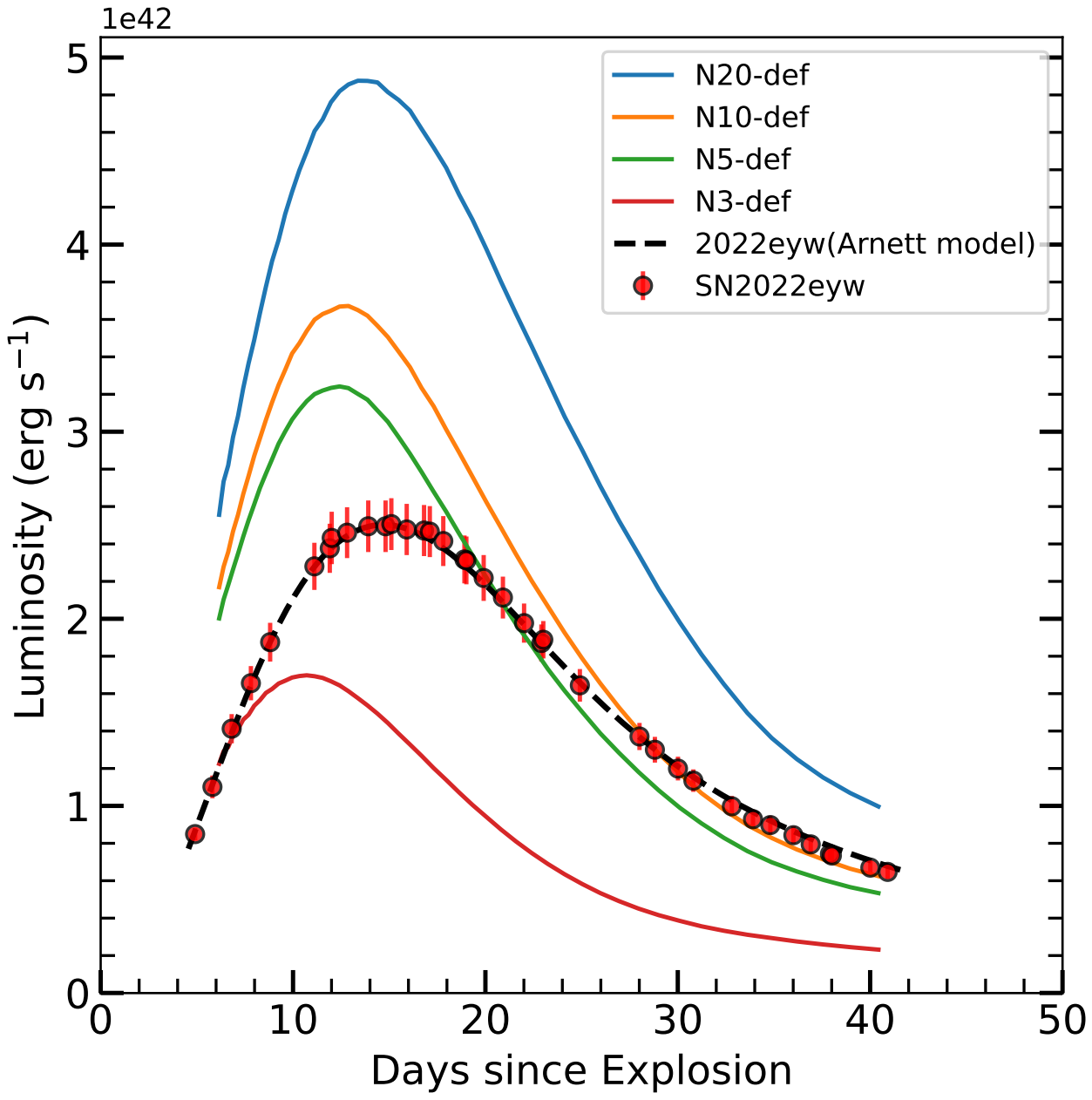

**Figure 6.** Comparison of the pseudobolometric light curve of SN 2022eyw (red circles) with synthetic bolometric light curves from the M. Fink et al. (2014) deflagration models. The models shown—N3-def, N5-def, N10-def, and N20-def—correspond to increasing numbers of ignition spots in the white dwarf and represent a sequence of increasing explosion energy and nickel mass. Also plotted is the one-dimensional radiation diffusion model fit with the observed pseudobolometric light curve.

This discrepancy is likely due to a mismatch in the ejecta mass predicted by the models and the actual ejecta mass of SN 2022eyw. This was also pointed out by M. R. Magee et al. (2016) for SN 2015H. The N3-def model, for example, has an ejecta mass of only $0.20\ M_\odot$, while for SN 2022eyw we estimate an ejecta mass of approximately $0.79\ M_\odot$ (refer to Section 4.3.2). A higher ejecta mass allows for more efficient $\gamma$-ray trapping, resulting in a longer photon diffusion timescale and thus a slower rise to the maximum and slower postmaximum decline in the light curve. Thus, even with similar $^{56}$Ni masses, the postmaximum evolution can differ significantly due to differences in total ejecta mass and therefore opacity (P. A. Pinto & R. G. Eastman 2000b).

We also examined whether variations in ignition geometry (e.g., the offset of the ignition spot) and white dwarf properties such as metallicity or central density could alleviate the discrepancies seen in the redder bands. F. Lach et al. (2022) investigated how these factors influence pure deflagration outcomes, and found out that the ignition offset and central density are the primary parameters controlling the $^{56}$Ni yield, while variations in metallicity or rotation have negligible effects. Their models use single-spot ignitions, which limit the burning efficiency, and even the brightest cases produce only modest $^{56}$Ni mass. They are named according to the ignition radius ("r10"), central density ("dX.X"), and metallicity ("Z"). For SN 2022eyw, we compared the observed multiband light curves with three of the brightest Lach models—r10_d4.0_Z, r10_d5.0_Z, and r10_d6.0_Z (Figure 7). None of these models reach the observed peak luminosity of SN 2022eyw. Similar to the Fink models, the redder bands of the Lach models decline too rapidly, reflecting insufficient ejecta mass and early transparency. This behavior is consistent with the conclusions of F. Lach et al. (2022), who showed that though the $^{56}$Ni mass is controlled by the central density and ignition radius, it is not enough to break the intrinsic correlation between ejecta mass and $^{56}$Ni mass in pure deflagrations. Consequently, these models cannot resolve the red-band discrepancy seen in SN 2022eyw. A comparison of peak magnitudes, $^{56}$Ni masses, ejecta masses, and kinetic energies for the Lach models and SN 2022eyw is provided in Table 3.

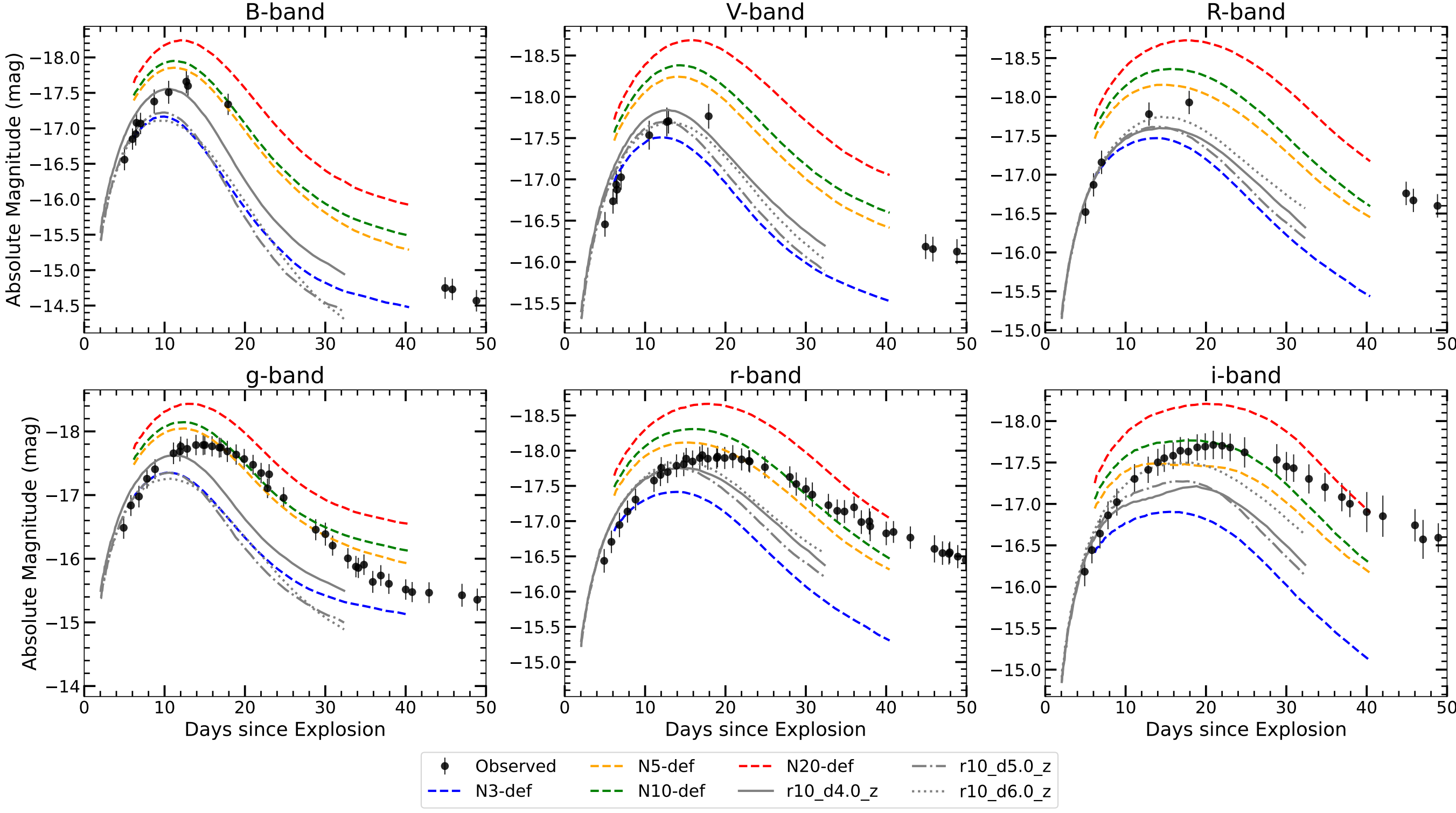

**Figure 7.** Comparison of observed *BgVRri* light curves of SN 2022eyw (data points) with synthetic photometry derived from the angle-averaged spectral time series of the M. Fink et al. (2014) deflagration models N3-def, N5-def, N10-def, and N20-def and the F. Lach et al. (2022) deflagration models r10_d4.0_Z, r10_d5.0_Z, and r10_d6.0_Z.

In the CO deflagration models, the $^{56}$Ni mass is tightly correlated with the ejecta mass because both are set by the extent of burning in the CO white dwarf. In SN 2022eyw, however, the inferred $^{56}$Ni mass is comparatively small for an ejecta mass of $\sim 0.79\,M_{\odot}$, suggesting that if a mechanism exists that can naturally suppress the $^{56}$Ni yield while still producing a relatively large ejecta mass, it may help resolve this mismatch. The hybrid CONe deflagration model of M. Kromer et al. (2015) illustrates a mechanism that naturally suppresses $^{56}$Ni production, as the flame quenches when it encounters the low-carbon ONe mantle. However, the model ejects only $\sim 3.4 \times 10^{-3}\,M_{\odot}$ of $^{56}$Ni and $\sim 0.014\,M_{\odot}$ in total—values characteristic of faint SNe Iax (e.g., SN 2008ha) and far below the $M_{\rm ej} \approx 0.79\,M_{\odot}$ and $M_{\rm Ni} \approx 0.11\,M_{\odot}$ estimated for SN 2022eyw. The extremely low ejecta mass also leads to rapid transparency and overly fast evolution in the redder bands, so this model cannot resolve the broad $R/r'/i'$-band behavior observed in SN 2022eyw. Thus, while it provides a physical pathway to reduce the Ni yield, the hybrid CONe scenario remains too weak to account for the luminosity and light-curve evolution of this event.

The observed red-band excess over models indicates the presence of additional sources of energy not included in the pure deflagration framework. M. Kromer et al. (2013) suggest the excess could be due to clumped iron-group ejecta, which can increase thermalization efficiency, or due to energy input by a bound remnant that remains after partial disruption of the white dwarf. Alternatively, interaction with a helium-rich companion or envelope may provide an additional energy source, as hypothesized for the luminous Type Iax events (C. McCully et al. 2022).

Besides CO or hybrid CONe deflagrations, two additional explosion pathways have been proposed for SNe Iax: CO–ONe white dwarf mergers and compact object–WD mergers. The CO–ONe merger simulations of R. Kashyap et al. (2018) eject only $\sim 0.08\,M_{\odot}$ with $\sim 10^{-3}\,M_{\odot}$ of $^{56}$Ni and produce very faint, rapidly evolving transients whose maximum brightness closely resembles that of the faintest Iax events, far below the luminosity of SN 2022eyw. Mergers involving a neutron star or black hole also yield low $^{56}$Ni masses and correspondingly

**Table 3**
Comparison of Multiband Light-curve Parameters, $^{56}$Ni Mass, Ejecta Mass ($M_{ej}$), and Kinetic Energy of the Ejecta ($E_k$) of SN 2022eyw and the Fink Deflagration Models

| Model | $\Delta m_{15}(B)$ | $\Delta m_{15}(g')$ | $\Delta m_{15}(r')$ | $M_B^{max}$ | $M_{g'}^{max}$ | $M_{r'}^{max}$ | $M_{Ni}$ | $M_{ej}$ | $E_k$ |
|---|---|---|---|---|---|---|---|---|---|
| SN 2022eyw | 1.46 | 1.43 | 0.62 | −17.61 | −17.80 | −17.90 | 0.11 | 0.79 | 1.93 |
| N3-def | 1.92 | 1.63 | 1.19 | −17.17 | −17.35 | −17.42 | 0.07 | 0.20 | 0.44 |
| N5-def | 1.69 | 1.47 | 0.96 | −17.85 | −18.05 | −18.12 | 0.16 | 0.37 | 1.35 |
| N10-def | 1.68 | 1.46 | 1.01 | −17.95 | −18.15 | −18.31 | 0.18 | 0.48 | 1.95 |
| N20-def | 1.56 | 1.37 | 0.94 | −18.24 | −18.43 | −18.66 | 0.26 | 0.86 | 3.75 |
| r10_d4.0_Z | 2.08 | 1.72 | 1.12 | −17.56 | −17.62 | −17.76 | 0.09 | 0.23 | 0.68 |
| r10_d5.0_Z | 2.22 | 1.87 | 1.24 | −17.22 | −17.35 | −17.75 | 0.08 | 0.24 | 0.75 |
| r10_d6.0_Z | 1.91 | 1.70 | 1.11 | −17.11 | −17.26 | −17.85 | 0.09 | 0.30 | 0.97 |

**Notes.** The $^{56}$Ni mass and ejecta mass are given in units of solar masses and the kinetic energy is given in units of $10^{50}$ erg s$^{-1}$. The decline rates $\Delta m_{15}$ and peak magnitudes ($M_\lambda$) are given for the $B$, $g'$, and $r'$ bands.

faint light curves, again comparable only to the faint Iax subclass (A. Bobrick et al. 2022).

While pure deflagration scenarios remain the most successful framework for explaining bright and intermediate-luminosity SNe Iax, the systematic discrepancies at both the bright end and faint end indicate that the current model grid is incomplete. Neither the hybrid CONe model nor the Lach models can decouple the tightly coupled relation between $^{56}$Ni mass and ejecta mass that determines both the peak luminosity and the width of the light curve. These limitations motivate further investigation into the systematic uncertainties in deflagration modeling, including radiative transfer assumptions, the role of a $^{56}$Ni-rich bound remnant, and the need for a broader exploration of multispot or more complex ignition geometries.

## 5. Spectral Analysis

### 5.1. Spectral Evolution

Figure 8 displays the spectral evolution of SN 2022eyw from −8.1 to +110.5 days since $B_{max}$. All phases are reported in rest-frame days relative to $B_{max}$. The earliest spectrum exhibits a hot, blue continuum, consistent with a high-temperature photosphere. Weak but discernible absorption features due to Fe III, Si III, and probable C III λ4647 (see Section 5 of A. Dutta et al. 2022) are evident between 4000 and 5000 Å. These high-ionization features are characteristic of brighter Type Iax events near peak (D. Branch et al. 2004; S. Jha et al. 2006) and suggest elevated ionization states in the outer ejecta. As the SN approaches maximum, the Fe III and Si III features strengthen, and additional lines appear, including S II, Si II λ6355, a Ca II near-IR (NIR) triplet, and possible C II λλ6580, 7234. The detection of carbon features, though tentative, indicates incomplete burning in the outer layers—consistent with a deflagration-driven explosion scenario (M. Kromer et al. 2013; M. Fink et al. 2014).

Near maximum light (−0.2 to +0.6 days), the Si II absorption remains relatively shallow. The evolution of the spectra shows a gradual shift from Fe III to Fe II dominance, accompanied by line blanketing in the blue. We have a long gap in spectral evolution between ∼+5 and ∼+30 days. During this period the spectrum evolves significantly. By +30.5 days, Fe II becomes the dominant species and the continuum becomes significantly redder. The Si and C features vanish and Fe II multiplets appear in place of them, as shown in Figure 8. The Ca II NIR triplet strengthens substantially, and absorption due to Cr II, Co II, and Na ID also becomes apparent (R. J. Foley et al. 2013). Between +30.5 and +71.0 days, the spectra evolve slowly, dominated by lines of IGEs, indicating the inner, metal-rich regions of the ejecta are now visible. Notably, the broadening and blending of Fe II and Co II lines remain moderate, consistent with low ejecta velocities (∼3500–5000 km s$^{-1}$). By +110.5 days, SN 2022eyw exhibits a mix of permitted and emerging forbidden lines, including [Fe II], [Ca II], and possibly [Ni II] near 7300 Å, suggesting that the spectrum is transitioning into the nebular phase but is not fully optically thin. This is a common trait of SNe Iax, which exhibit long-lived photosphere and delayed nebular transitions due to high central densities and fallback (S. Jha et al. 2006; R. J. Foley et al. 2016).

### 5.2. Spectral Comparison

The spectral sequence of SN 2022eyw was compared with other Type Iax events at similar phases, providing insights into different stages of its evolution relative to both brighter and fainter members of the subclass. For this purpose, we retrieved the comparison spectra from the Weizmann Interactive Supernova Data Repository (WISeREP;[8] O. Yaron & A. Gal-Yam 2012).

#### 5.2.1. Premaximum Phase

Figure 9 shows the premaximum spectrum of SN 2022eyw at −5.0 days, compared with those of other SNe Iax. The premaximum spectrum of SN 2022eyw is very similar to those of SN 2005hk, SN 2020rea, and SN 2020udy. They all show strong Fe III absorption near 4000–5000 Å, along with emerging Fe II lines and a weak Si II λ6355 feature, indicative of a hot photosphere. The intermediate-luminosity SN 2019muj shows a prominent C II absorption line, but it lacks strong Fe II and Si II features. The premaximum spectra of faint SN 2008ha and SN 2010ae show strong Si II and C II absorption features.

#### 5.2.2. Near-maximum Phase

The near-maximum spectrum (+0.6 days) of SN 2022eyw is compared with those of other well-studied SNe Iax, around a similar epoch, in Figure 10. The spectrum is remarkably similar to those of SN 2002cx, SN 2005hk, SN 2020rea, and SN 2020udy. All these SNe show strong Fe II and residual Fe III features, marking the cooling progression in the ejecta.

[8] https://www.wiserep.org/

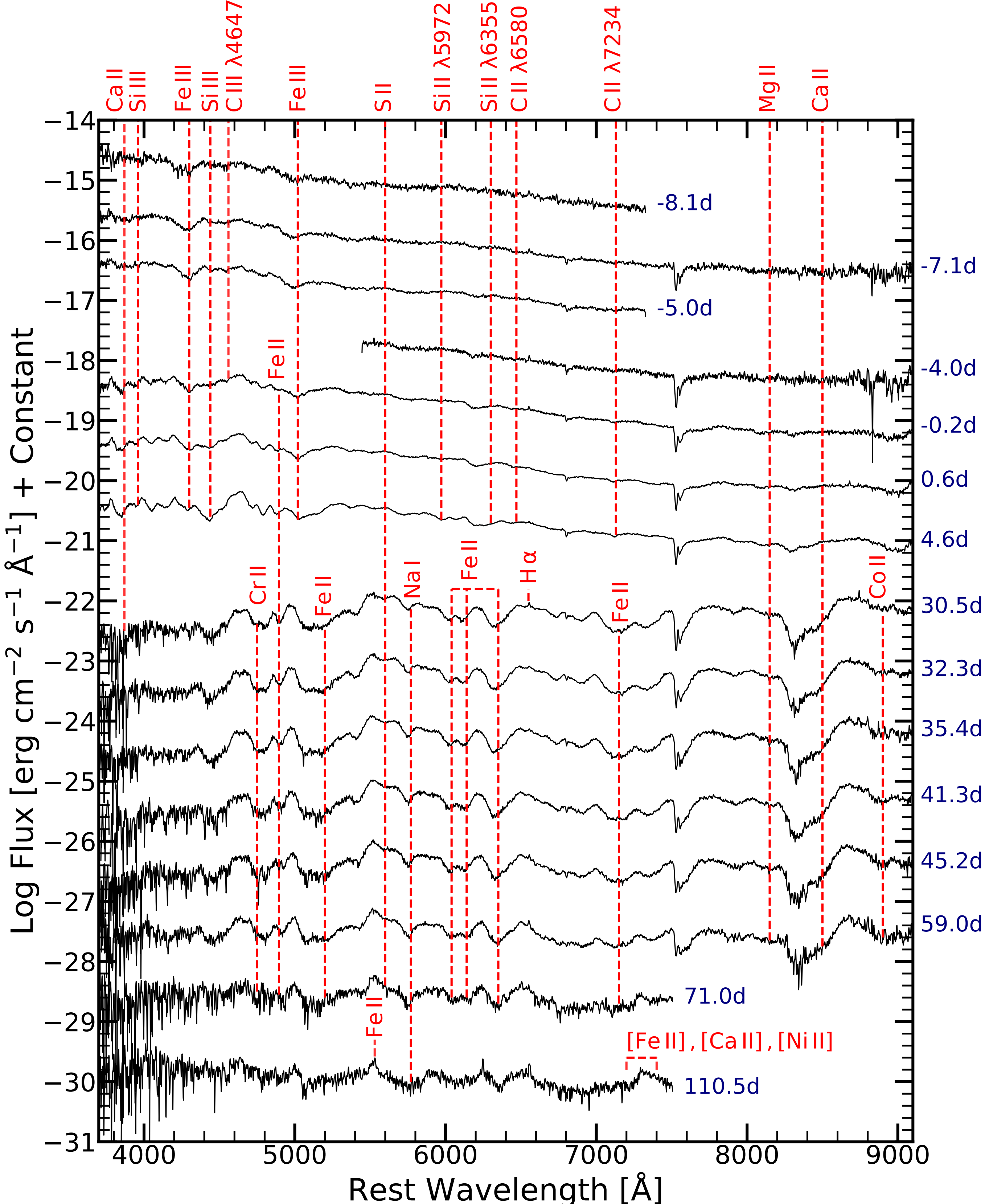

**Figure 8.** Spectral evolution of SN 2022eyw spanning −8.1 to +110.5 days relative to the *B*-band maximum. All spectra have been corrected for reddening and redshift, and smoothed to enhance clarity. Line identifications adopted are from D. K. Sahu et al. (2008) and A. Dutta et al. (2022).

(The data used to create this figure are available in the online article.)

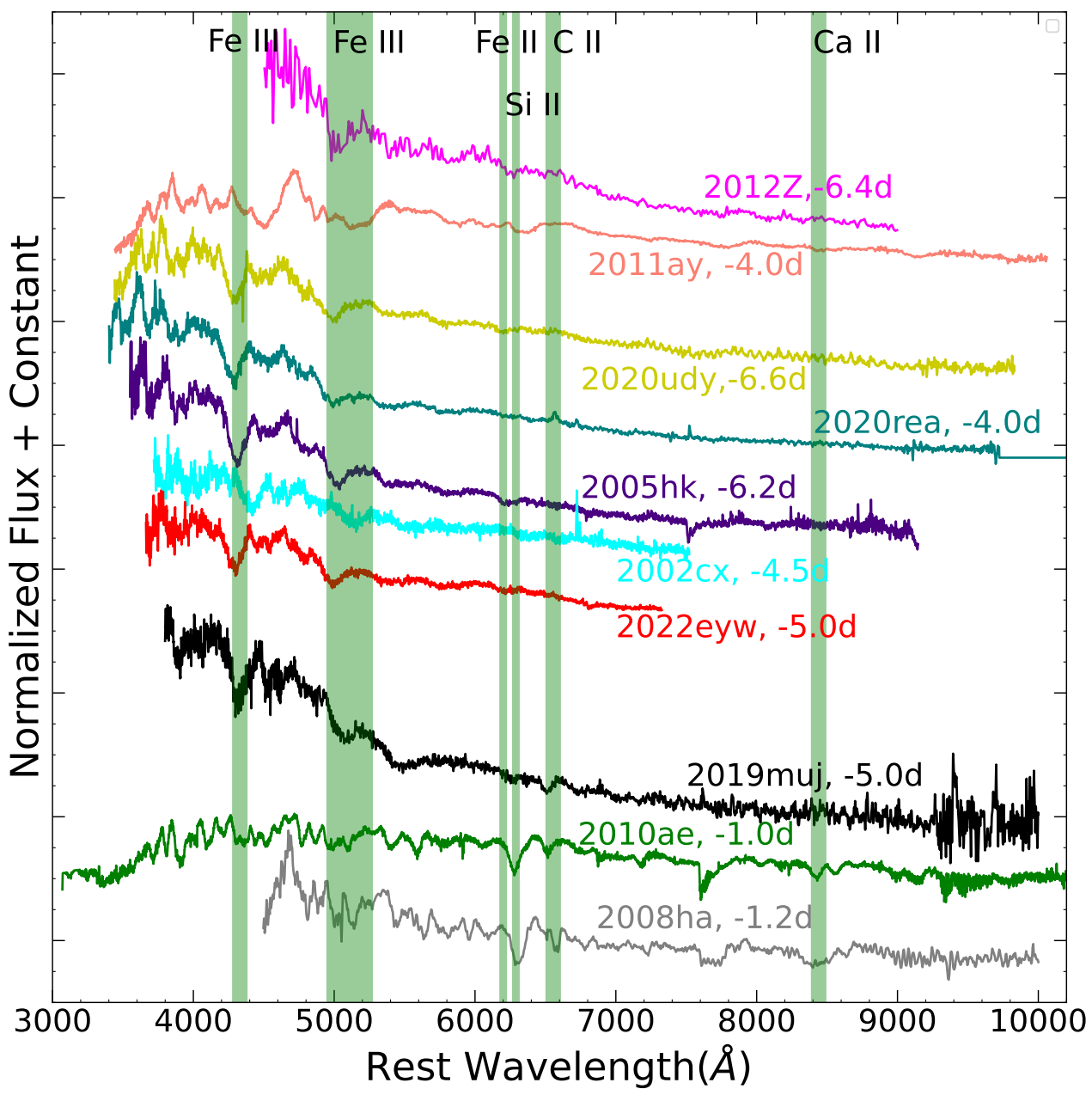

**Figure 9.** Premaximum spectrum of SN 2022eyw shown alongside those of other bright and faint SNe Iax at comparable phases. Spectra have been smoothed to enhance visual clarity.

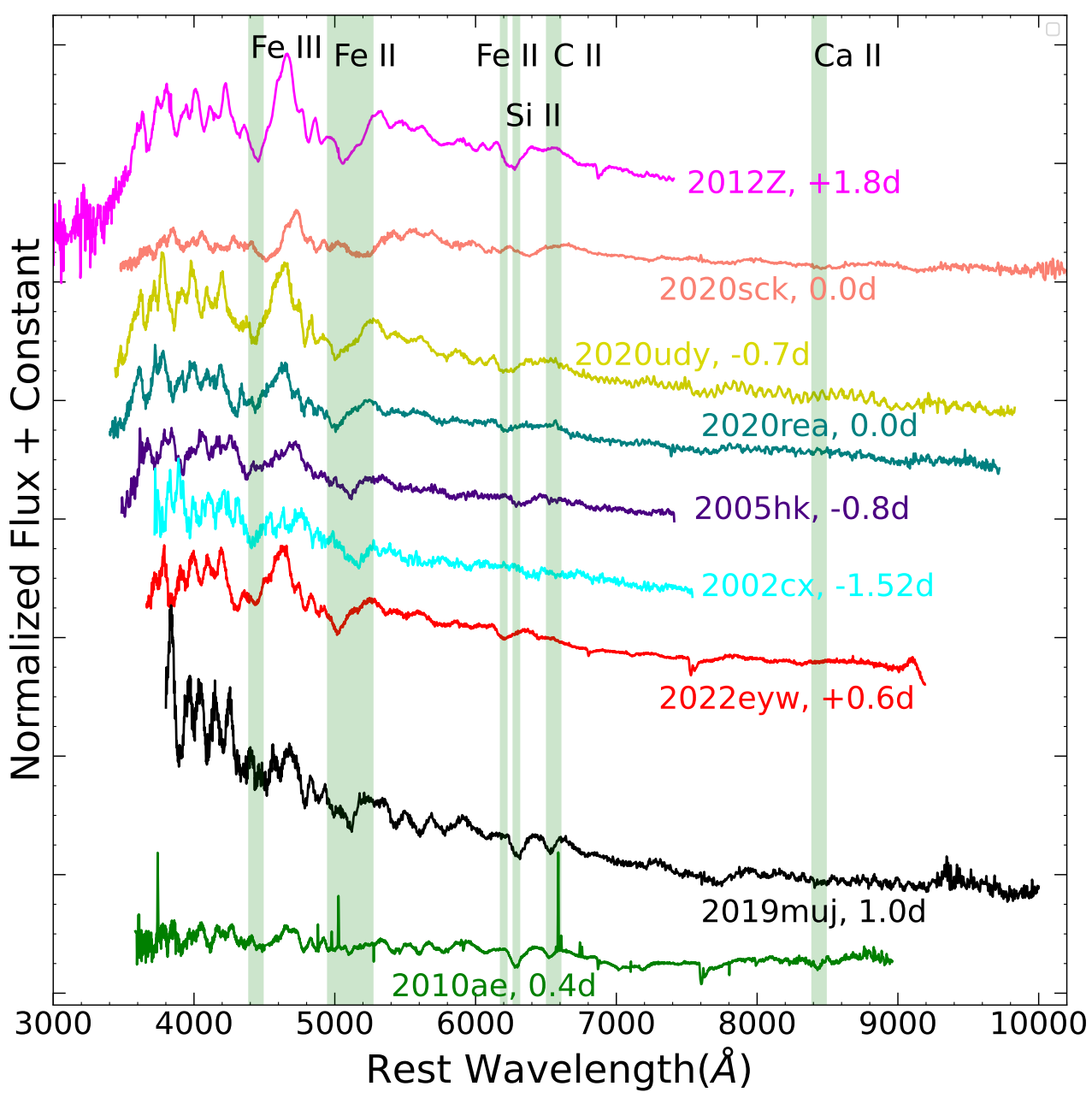

**Figure 10.** Comparison of SN 2022eyw's spectrum near maximum light with spectra of other well-observed Type Iax events. All spectra have been smoothed for clarity.

The blue region (4000–5000 Å) of the spectrum also shows features due to Ca II and Si III. A weak, broad Si II 6355 Å feature is also present. The spectrum differs from the intermediate-luminosity SN 2019muj and the fainter event SN 2010ae, which show stronger Si II, C II, and other intermediate-mass element (IME) lines (R. J. Foley et al. 2009; M. D. Stritzinger et al. 2014).

#### 5.2.3. Postmaximum Phase

Figure 11 shows the postmaximum spectrum of SN 2022eyw at +30.5 days, compared with those of other SNe Iax. By this phase, the spectrum is dominated by broad and blended Fe II features, marking the transition from a hotter Fe III–rich photosphere to a cooler, recombined state. The blue flux is significantly suppressed due to increased line blanketing, and the overall morphology closely resembles that of other bright Iax events such as SN 2002cx, SN 2005hk, SN 2012Z, SN 2020rea, and SN 2020udy. The lines due to Fe II multiplets, Cr II lines, and Co II lines near 9000 Å become prominent. The Ca II NIR triplet is clearly visible in the spectra at this phase. The postmaximum evolution of SN 2022eyw diverges from the intermediate-luminosity SN 2019muj and fainter SNe Iax, such as SN 2008ha and SN 2010ae, which show sharper and narrower spectral features due to a low expansion velocity of the ejecta.

#### 5.2.4. Late Phase

In Figure 12, the +110.5 day spectrum of SN 2022eyw is compared with the late-time spectra of other SNe Iax. Like its bright counterparts, SN 2022eyw shows a mix of permitted Fe II and Na ID lines and weak forbidden features like [Fe II], [Ca II], and [Ni II]. The absence of [Fe II] and [Fe III] features in the bluer regions indicates that SNe Iax retain their photospheres at late times (R. J. Foley et al. 2016) and do not completely go into the nebular phase.

The overall temporal evolution of SN 2022eyw places it firmly in the category of bright Type Iax events. Its spectral evolution—from hot, ionized early phases to a prolonged transition into the nebular regime—echoes the behavior of well-studied events like SN 2005hk, SN 2012Z, and SN 2020udy, reinforcing the interpretation of a low-energy, low-velocity explosion possibly with a surviving bound remnant.

### 5.3. Line Velocity Evolution

The photospheric velocity evolution of SN 2022eyw was estimated from the absorption minima of the Si II 6355 Å feature in the Doppler-corrected optical spectra. To determine the expansion velocity, Gaussian profiles were fitted to the Si II absorption troughs in the observed spectra. Reliable velocity measurements could be obtained up to +4.6 days with respect to $B_{\rm max}$, beyond which increasing line blending with emerging Fe lines makes it difficult to isolate the Si II component—a limitation also noted in earlier studies of SNe Iax (R. J. Foley et al. 2013; M. R. Magee et al. 2016). The error in velocity measurement is of the order of ∼150–450 km s$^{-1}$. The error possibly increases after the *B*-band maximum due to increased line blending effects. In Figure 13, the Si II velocity evolution of SN 2022eyw is plotted alongside a sample of other well-observed SNe Iax, including both luminous and fainter events.

At −7.1 days, SN 2022eyw exhibits a high expansion velocity of approximately 7000 km s$^{-1}$. This early-phase velocity is comparable to that of SN 2020rea and SN 2002cx. SN 2005hk shows slightly lower velocities at comparable premaximum epochs. While velocity measurements for faint SNe Iax such as SN 2008ha, SN 2019gsc, and SN 2010ae are unavailable at this early phase, their subsequent evolution suggests a distinctly lower velocity regime, with expansion speeds remaining below ∼4500 km s$^{-1}$ even near peak.

The velocity of the Si II $\lambda$6355 line in SN 2022eyw remains relatively flat in the premaximum phase, starting at

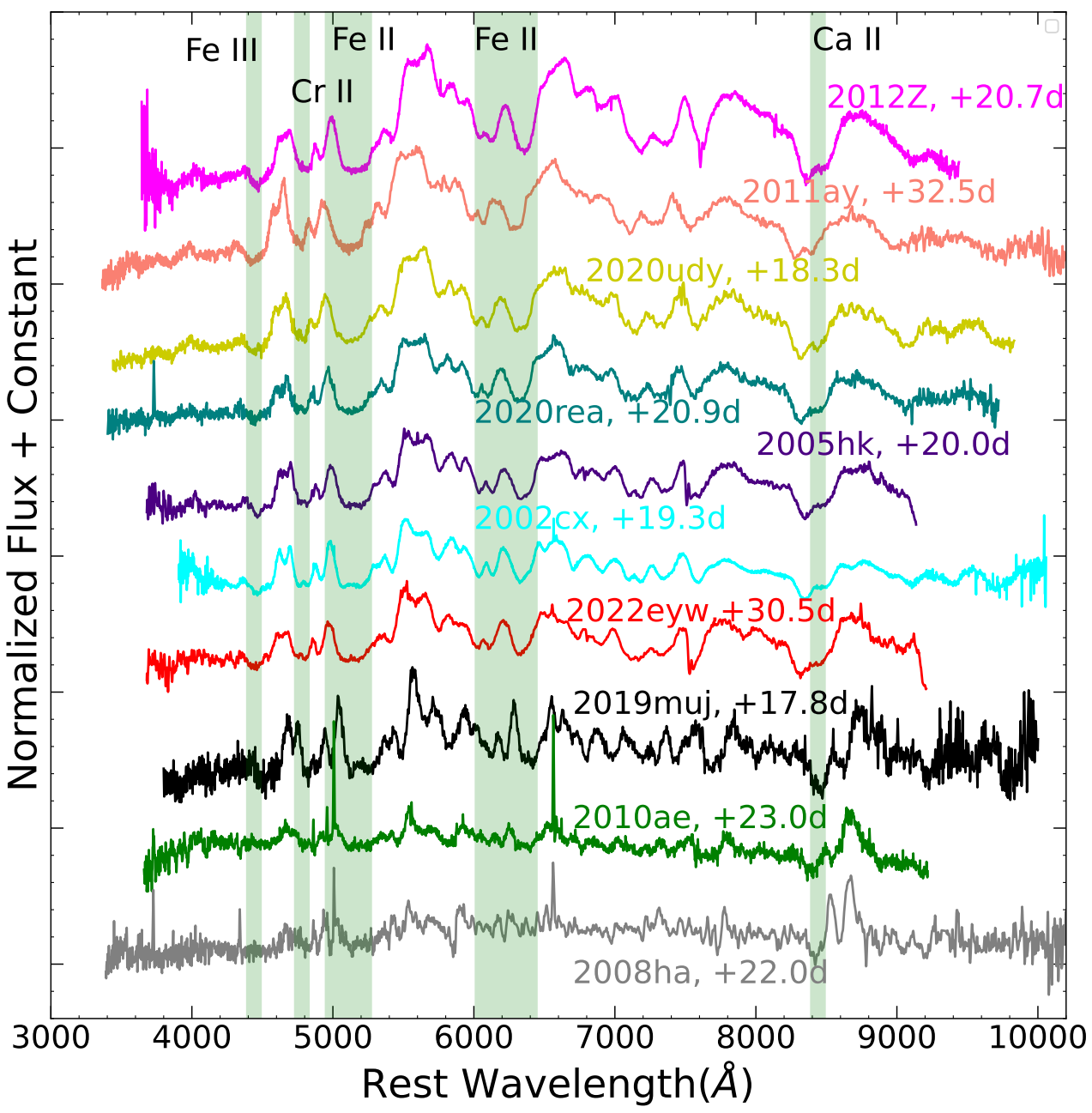

**Figure 11.** Spectral comparison of SN 2022eyw in the postmaximum phase with other bright and faint SNe Iax observed at similar epochs. Smoothed spectra are shown for ease of comparison.

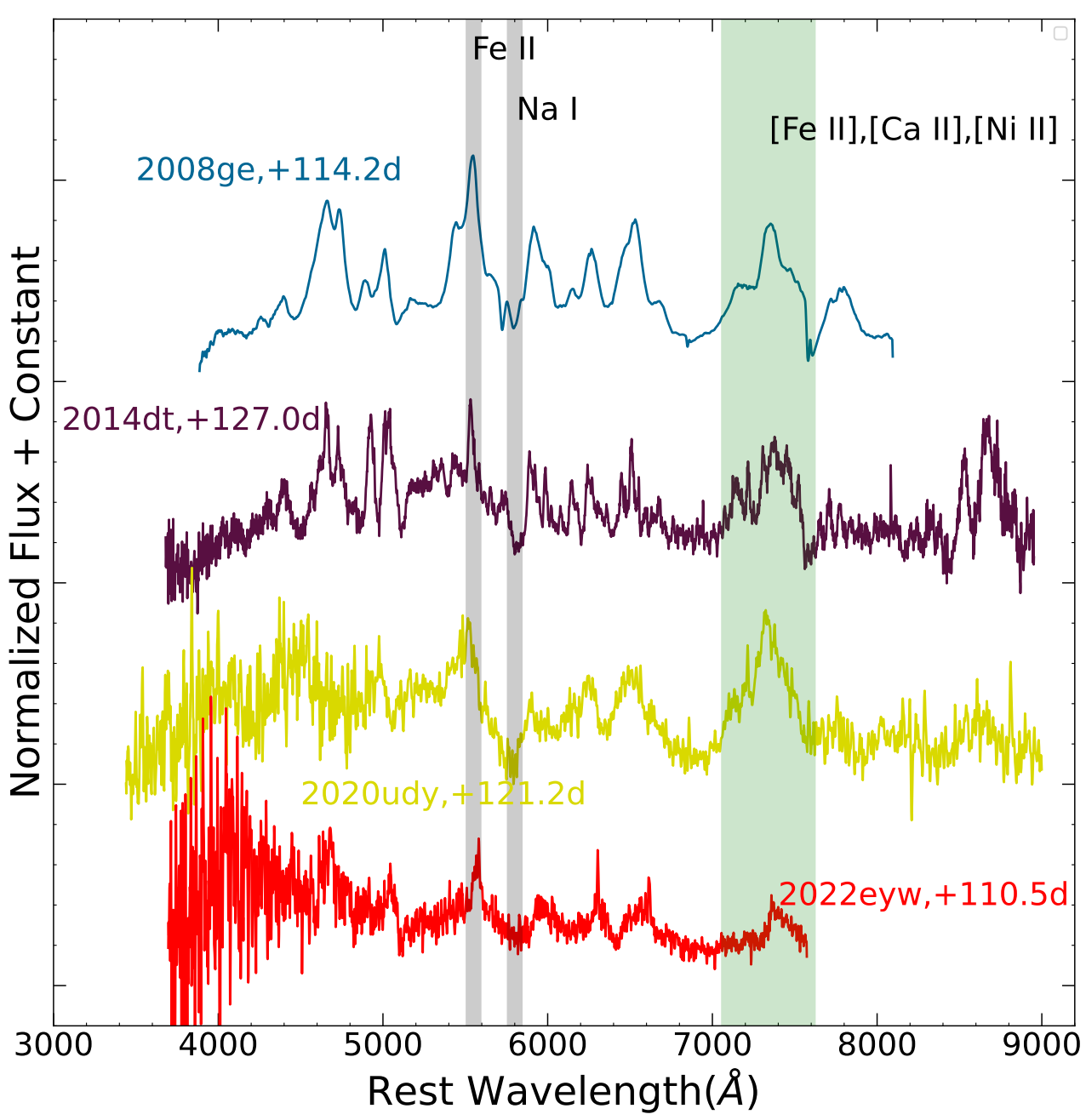

**Figure 12.** Late-time spectrum of SN 2022eyw compared to other SNe Iax representing a range of luminosities. All spectra have been smoothed.

$\sim$6900 km s$^{-1}$ at $-$5.0 days and only slightly declining to $\sim$6400 km s$^{-1}$ by $-$0.2 days. Around maximum light, however, the velocity begins to drop more rapidly, falling to around 5800 km s$^{-1}$ at +0.6 days and continuing to decline to $\sim$4800 km s$^{-1}$ by +4.6 days. It has to be noted here that in the red wing of Si II absorption, features due to Fe II start emerging, which could contaminate the Si II absorption feature. This may also lead to underestimation of Si II velocity if treated as a single, isolated line (M. Singh et al. 2025). Despite this, the overall trend is indicative of a retreating photosphere into deeper, slower-moving layers of the ejecta.

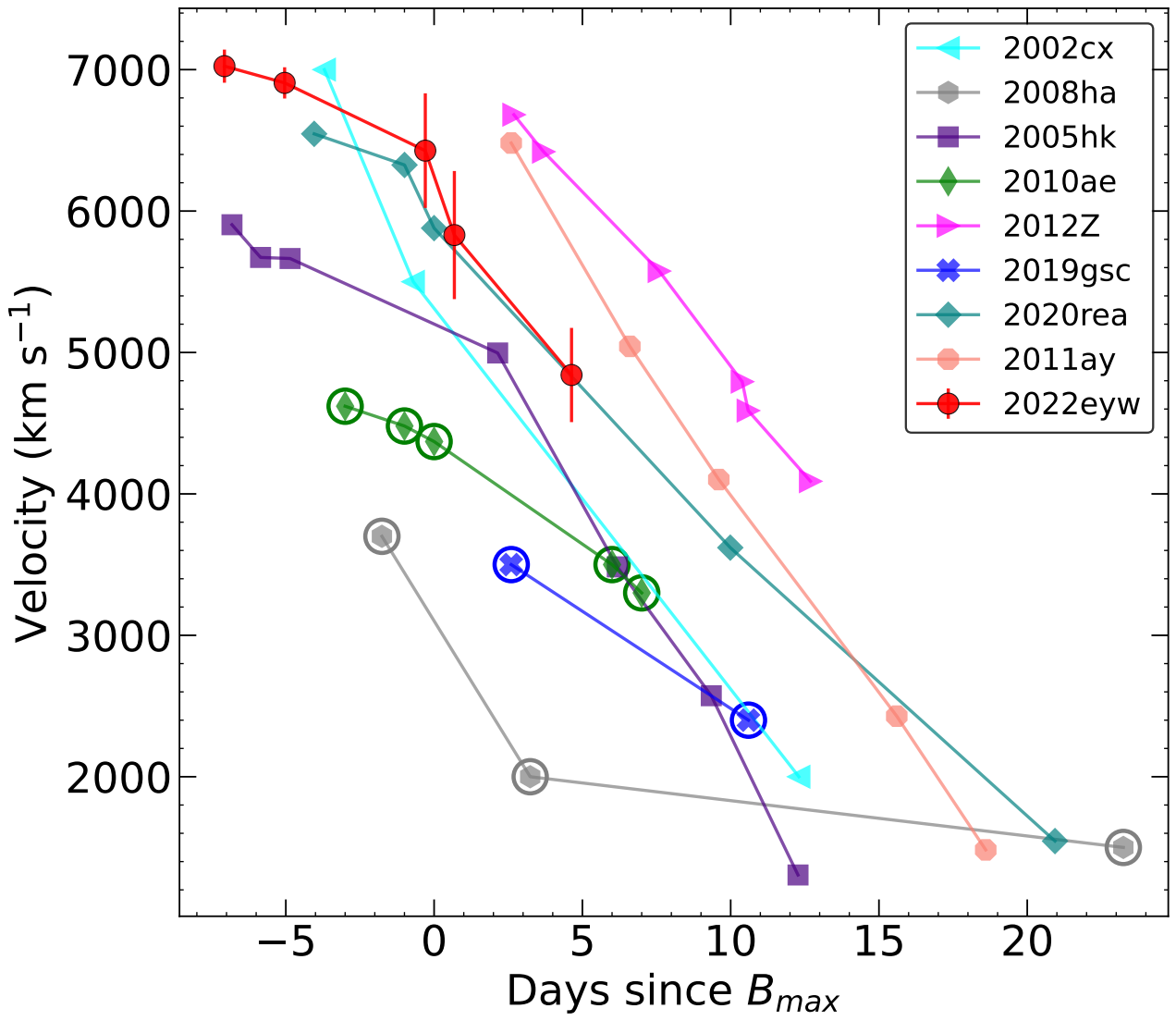

**Figure 13.** Evolution of the photospheric expansion velocity traced by the Si II $\lambda$6355 absorption feature for SN 2022eyw (red circles), compared with a sample of well-studied SNe Iax. Brighter SNe Iax in the comparison sample include SN 2002cx, SN 2005hk, SN 2011ay, SN 2012Z, SN 2020rea, and SN 2020udy, while fainter events such as SN 2008ha, SN 2010ae, and SN 2019gsc are highlighted using hollow circles. This plot illustrates the diversity in photospheric velocity evolution among SNe Iax and the convergence in velocities at later phases.

Compared to faint SNe Iax such as SN 2008ha, SN 2010ae, and SN 2019gsc, which exhibit significantly lower initial velocities due to their lower explosion energies, SN 2022eyw demonstrates higher expansion velocities characteristic of a more energetic explosion.

Interestingly, by around +10 days postmaximum, a convergence in expansion velocities is observed among both bright and faint SNe Iax. This convergence may be interpreted in terms of optical depth and ejecta structure. Due to the lower ejecta mass and kinetic energy in faint SNe Iax, the photosphere recedes more rapidly into slower-moving inner layers. Conversely, in bright Iax events with more massive ejecta, the outer faster layers obscure the inner material longer. However, by $\sim$10 days postmaximum, the outer ejecta have thinned sufficiently in both cases, and the observed photosphere has receded to slower-moving inner regions in all objects—naturally resulting in a convergence of measured velocities.

In addition to the Si II $\lambda$6355 velocity, we also measured the evolution of Fe III $\lambda$4420 and Fe III $\lambda$5156 line velocities in SN 2022eyw. At early epochs, the Fe III features exhibit the highest velocities, with Fe III $\lambda$4420 reaching over 10,000 km s$^{-1}$ at $-$8.1 days and remaining above 8000 km s$^{-1}$ until around maximum light. At the *B*-maximum, the velocities of Fe III $\lambda$4420 and Fe III $\lambda$5156 are around 8050 km s$^{-1}$ and 8170 km s$^{-1}$, respectively, while Si II $\lambda$6355 exhibits significantly lower velocities of $\sim$6400 km s$^{-1}$. Such a velocity distribution—where Fe III features appear at consistently higher velocities than those of Si II at similar phases—indicates the presence of both IGEs and IMEs over a wide velocity range in the ejecta. This suggests that the ejecta are not compositionally layered, but are consistent with a highly mixed explosion scenario (M. M. Phillips et al. 2007). Additional evidence for such mixing has also been reported for other bright SNe Iax, such as SN 2024pxl, where

**Table 4**
Fit Parameters of TARDIS Model

| Phase[a] | $v_{inner}$ | $L_{SN}$ | $X$(C) | $X$(O) | $X$(Ne) | $X$(Mg) | $X$(Si) | $X$(S) | $X$(Ca) | $X$(Ti) | $X$(Cr) | $X$(Fe) | $X$(Co) | $X$(Ni) |
|---|---|---|---|---|---|---|---|---|---|---|---|---|---|---|
| −5.0 | 6.8 | 8.8 | 0.0001 | 0.380 | 0.1856 | 0.001 | 0.0002 | 0.0001 | 0.005 | 0.00 | 0.00 | 0.011 | 0.007 | 0.410 |
| +0.7 | 5.9 | 8.95 | 0.001 | 0.180 | 0.207 | 0.100 | 0.032 | 0.008 | 0.005 | 0.005 | 0.020 | 0.030 | 0.002 | 0.410 |
| +4.6 | 5.0 | 8.9 | 0.001 | 0.120 | 0.218 | 0.100 | 0.032 | 0.008 | 0.005 | 0.005 | 0.020 | 0.080 | 0.001 | 0.410 |

**Note.**
[a] Time since *B*-band maximum (JD 2459672.55); $v_{inner}$: inner velocity of the ejecta ($10^3$ km s$^{-1}$); $v_{outer}$: outer velocity of the ejecta (fixed at 11,000 km s$^{-1}$); $L_{SN}$: luminosity of the SN (log $L_{\odot}$). The abundance of each species is denoted by $X$.

L. A. Kwok et al. (2025) observed centrally peaked emission features of both IGEs and IMEs with closely matching velocity widths and offsets. Such evidence of mixing supports the current hypothesis that pure deflagration of a near-Chandrasekhar-mass white dwarf is the leading explosion model for SNe Iax.

### 5.4. Spectral Modeling with TARDIS

To model the observed spectra of SN 2022eyw, we used the Monte Carlo radiative transfer code TARDIS (W. E. Kerzendorf & S. A. Sim 2014), which is widely used for generating synthetic spectra of SNe Ia during their photospheric phase. TARDIS simulates the propagation of energy packets representing photons through a spherically symmetric and homologously expanding ejecta, assuming an inner boundary emitting a blackbody continuum. The code traces the interactions of these energy packets with the material above the inner boundary to compute the emergent spectrum.

To generate a synthetic spectrum, TARDIS requires several key input parameters: the luminosity of the SN ($L_{SN}$ in terms of $\log L_{\odot}$), the time since explosion ($t_{exp}$ in days), a density profile, and the mass fractions of the elements. The model ejecta above the inner boundary (denoted by $v_{inner}$ in velocity space) is divided into spherically symmetric shells, and the code iteratively solves for the plasma parameters like the radiation temperature ($T_{rad}$) and dilution factor ($W$). Spectral modeling with TARDIS is particularly useful for diagnosing the composition and structure of the ejecta. SNe Iax are believed to result from pure deflagration leading to significant mixing in the ejecta (D. Branch et al. 2004; M. M. Phillips et al. 2007). As a result, uniform mass fractions were adopted in our models, rather than a stratified one, to incorporate the expected macroscopic mixing and lack of strong compositional layering. However, it should be noted that there could possibly be inhomogeneities in the composition in some directions of the explosion. M. Fink et al. (2014) studied deflagration in CO WDs and their angle-averaged models are to a large extent uniform throughout the ejecta. We used an exponential density profile of the form

$$\rho(v,\, t_{exp}) = \rho_0 e^{-\frac{v}{v_{exp}}} \left(\frac{t_{exp}}{t_0}\right)^{-3} \tag{5}$$

where the density becomes $\frac{1}{e}$ of $\rho_0$ at a velocity $v_{exp}$, and $t_0$. The values of $v_{exp}$, $\rho_0$, and $t_0$ used in the simulations were 7000 km s$^{-1}$, $8 \times 10^{-11}$ g cm$^{-3}$, and 2 days, respectively. The outer velocity ($v_{outer}$) was kept fixed at 11,000 km s$^{-1}$. For modeling the spectra, the inner velocity was decreased from 6800 km s$^{-1}$ at $\sim$8.0 days to 5000 km s$^{-1}$ at $\sim$17.8 days since explosion. In TARDIS, the features that contribute to the model spectra are formed above an inner velocity ($v_{inner}$). As time increases, this inner velocity decreases, and the region of the ejecta which contributes to the spectral features now recedes inward. Thus, for each epoch, the line-forming region is different, and although we assumed uniform abundances within that region, we adjusted the mass fractions between epochs to get a better fit (in a $\chi$-by-eye sense) to the observations (see Section 6.3.1 of D. K. Sahu et al. 2008). However, it is important to note that, except for the isotopic mass fractions, the composition of other elements does not change in the SN ejecta after it has reached homologous expansion.

For this study, we modeled synthetic spectra at three representative epochs: a premaximum phase (−5.0 days), near-maximum light (0.6 days), and a postmaximum phase (4.6 days). At each of these epochs, we adjusted the model luminosity, inner boundary velocity, and elemental mass fractions to achieve a better match with the observed spectra. To compare the observed spectra with the synthetic spectra produced by TARDIS, we converted the observed spectra from flux scale to luminosity density scale by adopting a distance of $40.6 \pm 2.8$ Mpc, estimated using the Virgo infall-corrected velocity (see Section 3.1). We list the TARDIS settings and mass fractions for each epoch in Table 4.

At the earliest epoch (−5.0 days), the ejecta is dominated by unburnt oxygen and radioactive $^{56}$Ni, with minimal contributions from IMEs like Mg and Si, and IGEs such as Fe and Co. This composition reflects the outer, less processed layers of the ejecta. By maximum light (+0.6 days), deeper regions of the ejecta contribute to the spectrum. Here, we require significantly enhanced abundances of IMEs, particularly Mg and Si, as well as a noticeable increase in Fe, Cr, and Ti, suggesting these elements reside at intermediate depths. At the postmaximum epoch (+4.6 days), the inner ejecta layers become visible, requiring an even higher Fe mass fraction (0.08) and sustained levels of IMEs. These trends align with expectations from pure deflagration models that predict strong mixing and shallow abundance gradients in SNe Iax.

Traces of carbon are present in all three epochs, with the earliest spectrum (−5.0 days) showing a faint but identifiable feature around 4600 Å, consistent with C III $\lambda$4647. Our model successfully reproduces this feature, which has also been reported in other SNe Iax, including SN 2014ck (L. Tomasella et al. 2016) and SN 2020sck (A. Dutta et al. 2022). The persistent, though low, carbon abundance across epochs may point to residual unburnt material from the progenitor. Although oxygen is not directly identified in the observed spectra, its lack of visibility may result from line blanketing by Fe II features forming at comparable or higher velocities, effectively masking potential signatures of unburnt oxygen in the outer ejecta (E. Baron et al. 2003). The presence of these light elements—especially carbon and oxygen—in combination with a dominant $^{56}$Ni component, supports the possibility of incomplete burning and further strengthens the hypothesis of a pure deflagration in a CO white dwarf progenitor, as proposed in theoretical studies (e.g., M. Kromer et al. 2013).

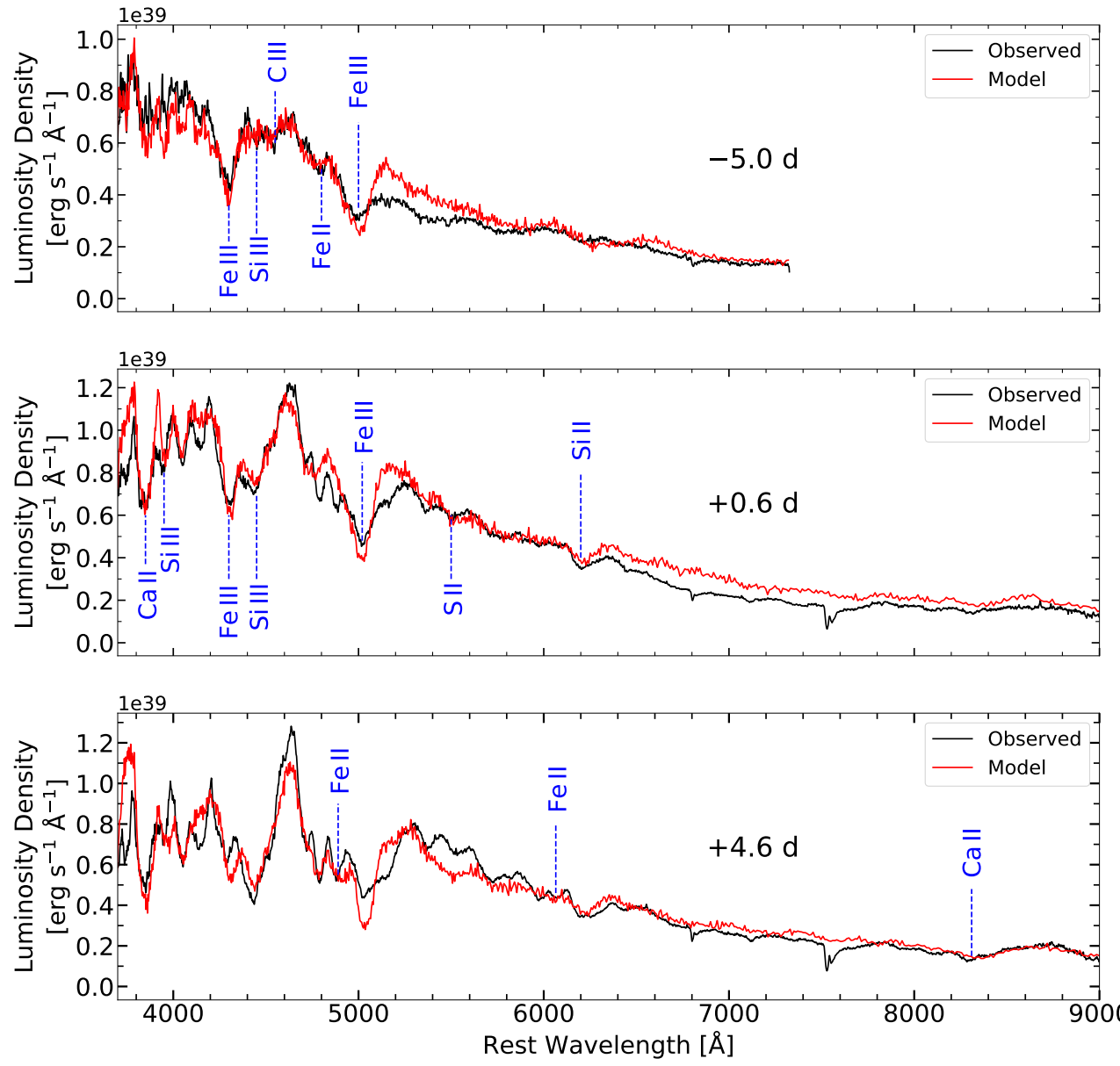

**Figure 14.** Comparison of the observed spectra of SN 2022eyw with synthetic spectra generated using `TARDIS` at three different epochs: −5.0, +0.6, and +4.6 days relative to $B$-band maximum. The observed spectra (black) are shown alongside the corresponding `TARDIS` model spectra (red). Prominent spectral features are marked, and phases are indicated in each panel.

Our fitting suggests that a uniform composition of elements performs reasonably well to reproduce the prominent features in the spectra (see Figure 14). However, a detailed exploration of the fitting parameters is beyond the scope of this work.

## 6. Summary

A detailed study of the bright Type Iax SN 2022eyw based on photometric and spectroscopic observations is presented in this work. SN 2022eyw reached a peak absolute magnitude of $M_B = -17.61 \pm 0.15$ mag with a rise time of ∼13 days, and $M_g = -17.80 \pm 0.15$ mag with a rise time of ∼15 days. The decline rates were measured as $\Delta m_{15}(B) = 1.46 \pm 0.05$ and $\Delta m_{15}(g') = 1.43 \pm 0.06$. These values place SN 2022eyw among the brighter members of the Iax class, showing a close resemblance to SNe Iax SN 2005hk, SN 2012Z, SN 2020rea, and SN 2020udy, and thereby increasing the statistical sample of bright SNe Iax.

Modeling of the pseudobolometric light curve yields a synthesized $^{56}$Ni mass of $M_{\rm Ni} = 0.11^{+0.01}_{-0.01}\, M_\odot$. A photospheric velocity of ∼6400 km s$^{-1}$ near maximum yields an ejecta mass of $0.79^{+0.08}_{-0.07}\, M_\odot$ and a kinetic energy of explosion of $0.19^{+0.02}_{-0.01} \times 10^{51}$ erg. A comparison of the light curve with current pure deflagration models places SN 2022eyw between the N3-def and N5-def cases of Fink's model, supporting the interpretation that SN 2022eyw arose from a low-energy, partial deflagration event. However, discrepancies with the models, such as its slower postmaximum decline and excess red flux, suggest that present simulations do not yet fully capture the observations.

The spectral evolution from −8 to +110 days post $B$-band maximum exhibits the typical features of bright SNe Iax, with Fe III and Si III features during the early evolution giving way to Fe II and Co II dominance at later phases. `TARDIS` modeling of the early spectra indicates well-mixed, Fe group–dominated ejecta with traces of unburnt carbon, consistent with incomplete burning in pure deflagration models.

The similarities of SN 2022eyw to other well-studied events support the idea that bright SNe Iax constitute a relatively homogeneous subset within the broader Iax population, yet the variations in velocity evolution and light-curve behavior indicate that ignition geometry, flame strength, and radiative transfer effects must all play a role in shaping their diversity. Moreover, the evidence for incomplete burning and relatively low kinetic energy leaves open the possibility that SN 2022eyw, like other members of its class, may have left behind a bound remnant, offering a valuable probe of failed or partial thermonuclear disruption. Taken together, the results presented here highlight partial deflagration of a Chandrasekhar-mass CO white dwarf as the most promising framework for explaining bright SNe Iax, while also underscoring the limitations of current model predictions.

In addition to its overall similarity to other bright events, SN 2022eyw also provides several new observational constraints at the luminous end of the Type Iax population. Its high ejecta mass, together with a comparatively modest $^{56}$Ni yield, highlights a tension with current pure deflagration models, which typically predict a stronger correlation between these quantities. Additionally, our comparison with hybrid CONe models, white dwarf merger scenarios, and compact-object channels shows that these alternatives cannot reproduce the observed luminosity and ejecta mass combination. Consequently, SN 2022eyw strengthens the evidence for a near-Chandrasekhar explosion origin for bright SNe Iax while also motivating more detailed parameter studies within the deflagration framework. Future progress will require multidimensional explosion simulations with improved radiative transfer treatments, coupled with well-sampled multiwavelength observations, to place stronger constraints on the physical origins and remnants of these peculiar explosions.

## Acknowledgments

This work is carried out as part of the project "Understanding the progenitor and explosion mechanism of supernovae through studies in the nearby Universe" (CRG/2022/007688). Funding for this project by the Science and Engineering Research Board (SERB), Department of Science and Technology, Government of India, is acknowledged.

We thank the staff of the Indian Astronomical Observatory (IAO), Hanle, and the CREST campus, Hosakote, for their assistance in making observations possible with the Himalayan Chandra Telescope (HCT). This work has also benefited from data obtained with the GROWTH-India Telescope (GIT) set up by the Indian Institute of Astrophysics (IIA) and the Indian Institute of Technology Bombay (IITB) with funding from the Indo-US Science and Technology Forum and the Science and Engineering Research Board, Department of Science and Technology, Government of India. It is located at the IAO. We acknowledge funding by the IITB alumni batch of 1994, which partially supports operations of the telescope. The observational facilities at IAO and CREST are operated by IIA, Bengaluru, an autonomous institute under the Department of Science and Technology, Government of India. We also thank the observers of HCT and GIT for carrying out target-of-opportunity (ToO) observations during the early follow-up campaign.

H.D. and D.K.S. acknowledge support from the CRG/2022/007688 project. G.C.A. acknowledges support from the Indian National Science Academy (INSA) under its Senior Scientist Programme. A.D. acknowledges funding support from HST-AR-16613.002-A and HST-GO-16885.011-A. M.S. acknowledges financial support provided under the National Post Doctoral Fellowship (N-PDF; file number: PDF/2023/002244) by the Science & Engineering Research Board (SERB), Anusandhan National Research Foundation (ANRF), Government of India.

This work has made use of data and tools provided by the NASA Astrophysics Data System[9] (ADS), and the NASA/IPAC Extragalactic Database (NED). We acknowledge the Weizmann Interactive Supernova Data Repository (WISeREP). Photometric data from the Swift/UVOT archive and the Zwicky Transient Facility (ZTF) archive accessed via Lasair[10] (K. W. Smith et al. 2019) have also been used in this study.

For completeness, the full *UBVR* and *griz* photometric measurements of SN 2022eyw are provided in Tables 5 and 6. Details of the spectroscopic follow-up are given in Table 7.

The analysis made extensive use of a community-developed software package, TARDIS v2023.11.26, for spectral synthesis in supernovae (W. Kerzendorf et al. 2018, 2019). Its development was supported by the Google Summer of Code and ESA's Summer of Code in Space programs. TARDIS builds on several packages, including Astropy (Astropy Collaboration et al. 2013) and PyNE (A. Scopatz et al. 2012). The simulations employed the TARDIS atomic data file named kurucz_cd23_chianti_H_He.h5 (HDF5 format) with UUID: 6f7b09e887a311e7a06b246e96350010 and MD5: 864f1753714343c41f99cb065710cace. We also made use of the Heidelberg Supernova Model Archive[11] (HESMA; M. Kromer et al. 2017). Additional software tools utilized in this work include (i) IRAF (D. Tody 1993), (ii) PyRAF (Science Software Branch at STScI 2012), (iii) NumPy (S. Van Der Walt et al. 2011), (iv) Matplotlib (J. D. Hunter 2007), (v) SciPy (P. Virtanen et al. 2020), (vi) pandas (W. Mckinney 2011), (vii) Astropy (Astropy Collaboration et al. 2013, 2018, 2022), (viii) emcee (D. Foreman-Mackey et al. 2013), (ix) corner (D. Foreman-Mackey 2016), (x) sncosmo (K. Barbary et al. 2016), and (xi) extinction (K. Barbary 2021).

## Appendix

Table 5 presents the photometric data of SN 2022eyw in *UBVR* bands, while the photometric data in *griz* bands are listed in Table 6. The log of the spectroscopic observations of SN 2022eyw, including details of the observing epochs and wavelength coverage, is provided in Table 7.

**Table 5**
Photometry of SN 2022eyw in *UBVR* Bands (Vega System)

| Date | JD[a] | Phase[b] | *U* | *B* | *V* | *R* | Telescope |
|---|---|---|---|---|---|---|---|
| 2022-03-25 | 664.4 | −8.2 | ... | 16.76 ± 0.007 | 16.80 ± 0.004 | 16.69 ± 0.007 | HCT |
| 2022-03-26 | 665.4 | −7.2 | 15.63 ± 0.04 | 16.47 ± 0.005 | 16.52 ± 0.003 | 16.34 ± 0.006 | HCT |
| 2022-03-27 | 665.8 | −6.8 | 15.52 ± 0.06 | 16.40 ± 0.06 | 16.32 ± 0.09 | ... | Swift/UVOT |
| 2022-03-27 | 665.9 | −6.7 | 15.50 ± 0.06 | 16.24 ± 0.06 | 16.38 ± 0.09 | ... | Swift/UVOT |
| 2022-03-27 | 666.4 | −6.2 | 15.48 ± 0.03 | 16.25 ± 0.015 | 16.23 ± 0.014 | 16.05 ± 0.019 | HCT |
| 2022-03-29 | 668.1 | −4.5 | 15.29 ± 0.08 | 15.94 ± 0.08 | ... | ... | Swift/UVOT |
| 2022-03-31 | 669.9 | −2.7 | ... | 15.81 ± 0.06 | 15.72 ± 0.09 | ... | Swift/UVOT |
| 2022-04-02 | 672.1 | −0.5 | ... | 15.66 ± 0.07 | 15.56 ± 0.09 | ... | Swift/UVOT |
| 2022-04-02 | 672.3 | −0.3 | 15.21 ± 0.02 | 15.72 ± 0.008 | 15.55 ± 0.003 | 15.43 ± 0.005 | HCT |
| 2022-04-07 | 677.3 | 4.8 | 15.48 ± 0.17 | 15.98 ± 0.007 | 15.49 ± 0.006 | 15.28 ± 0.007 | HCT |
| 2022-05-04 | 704.3 | 31.8 | ... | 18.57 ± 0.027 | 17.07 ± 0.02 | 16.45 ± 0.021 | HCT |
| 2022-05-05 | 705.2 | 32.7 | 18.79 ± 0.04 | 18.59 ± 0.019 | 17.10 ± 0.005 | 16.54 ± 0.01 | HCT |
| 2022-05-08 | 708.2 | 35.7 | ... | 18.75 ± 0.026 | 17.13 ± 0.018 | 16.61 ± 0.019 | HCT |
| 2022-05-14 | 714.2 | 41.7 | ... | 18.77 ± 0.027 | 17.33 ± 0.007 | 16.79 ± 0.012 | HCT |
| 2022-05-18 | 718.3 | 45.8 | ... | 18.84 ± 0.022 | 17.36 ± 0.01 | 17.00 ± 0.015 | HCT |
| 2022-06-01 | 732.2 | 59.7 | 19.14 ± 0.11 | ... | 17.75 ± 0.014 | 17.22 ± 0.022 | HCT |

**Notes.** Missing data are denoted by ellipses.
[a] JD offset = JD − 2459000.
[b] Phase in days relative to *B*-band maximum at JD = 2459672.55.
(This table is available in its entirety in machine-readable form in the online article.)

[9] https://ui.adsabs.harvard.edu/
[10] https://lasair-lsst.lsst.ac.uk/
[11] https://hesma.h-its.org

**Table 6**
Photometry of SN 2022eyw in *griz* Bands (AB System)

| Date | JD[a] | Phase[b] | *g* | *r* | *i* | *z* | Telescope |
|---|---|---|---|---|---|---|---|
| 2022-03-25 | 664.3 | −9.7 | 16.81 ± 0.09 | 16.78 ± 0.08 | 16.98 ± 0.09 | 17.28 ± 0.11 | GIT |
| 2022-03-26 | 665.2 | −8.8 | 16.46 ± 0.07 | 16.51 ± 0.06 | 16.72 ± 0.05 | 16.93 ± 0.08 | GIT |
| 2022-03-27 | 666.2 | −7.8 | 16.32 ± 0.08 | 16.27 ± 0.09 | 16.52 ± 0.10 | 16.64 ± 0.12 | GIT |
| 2022-03-28 | 667.2 | −6.8 | 16.04 ± 0.04 | 16.08 ± 0.05 | 16.30 ± 0.08 | 16.40 ± 0.15 | GIT |
| 2022-03-29 | 668.2 | −5.8 | 15.89 ± 0.04 | 15.91 ± 0.03 | ... | 16.35 ± 0.08 | GIT |
| 2022-03-29 | 668.3 | −5.7 | ... | ... | 16.14 ± 0.06 | ... | GIT |
| 2022-04-01 | 670.5 | −3.5 | 15.64 ± 0.08 | 15.64 ± 0.06 | 15.86 ± 0.10 | 16.00 ± 0.11 | GIT |
| 2022-04-01 | 671.3 | −2.7 | 15.61 ± 0.05 | 15.56 ± 0.05 | ... | 15.99 ± 0.08 | GIT |
| 2022-04-01 | 671.4 | −2.6 | 15.53 ± 0.03 | 15.46 ± 0.03 | ... | ... | ZTF |
| 2022-04-02 | 672.2 | −1.8 | 15.57 ± 0.06 | 15.52 ± 0.04 | 15.75 ± 0.05 | 15.93 ± 0.12 | GIT |
| 2022-04-03 | 673.3 | −0.7 | 15.51 ± 0.06 | 15.43 ± 0.06 | ... | ... | GIT |
| 2022-04-03 | 673.4 | −0.6 | ... | ... | 15.66 ± 0.07 | 15.87 ± 0.08 | GIT |
| 2022-04-04 | 674.2 | 0.2 | 15.51 ± 0.05 | 15.41 ± 0.08 | 15.61 ± 0.07 | 15.77 ± 0.11 | GIT |
| 2022-04-04 | 674.4 | 0.4 | 15.51 ± 0.02 | ... | ... | ... | ZTF |
| 2022-04-05 | 674.5 | 0.5 | ... | 15.33 ± 0.03 | ... | ... | ZTF |
| 2022-04-05 | 675.3 | 1.3 | 15.53 ± 0.05 | 15.37 ± 0.05 | 15.58 ± 0.06 | 15.70 ± 0.07 | GIT |
| 2022-04-06 | 676.2 | 2.2 | 15.55 ± 0.05 | 15.32 ± 0.06 | 15.52 ± 0.07 | 15.65 ± 0.11 | GIT |
| 2022-04-06 | 676.4 | 2.4 | 15.55 ± 0.02 | ... | ... | ... | ZTF |
| 2022-04-07 | 676.5 | 2.5 | ... | 15.28 ± 0.03 | ... | ... | ZTF |
| 2022-04-07 | 677.2 | 3.2 | 15.61 ± 0.07 | 15.33 ± 0.05 | 15.53 ± 0.06 | 15.58 ± 0.09 | GIT |
| 2022-04-08 | 678.3 | 4.3 | 15.66 ± 0.06 | 15.32 ± 0.04 | 15.48 ± 0.05 | 15.61 ± 0.12 | GIT |
| 2022-04-08 | 678.4 | 4.4 | ... | 15.30 ± 0.03 | ... | ... | ZTF |
| 2022-04-09 | 679.3 | 5.3 | 15.73 ± 0.05 | 15.32 ± 0.05 | 15.47 ± 0.06 | 15.53 ± 0.10 | GIT |
| 2022-04-10 | 680.3 | 6.3 | ... | 15.30 ± 0.06 | 15.45 ± 0.08 | 15.57 ± 0.10 | GIT |
| 2022-04-10 | 680.4 | 6.4 | 15.82 ± 0.04 | ... | ... | ... | GIT |
| 2022-04-11 | 681.4 | 7.4 | 15.95 ± 0.06 | 15.34 ± 0.05 | 15.46 ± 0.05 | 15.59 ± 0.11 | GIT |
| 2022-04-12 | 682.2 | 8.2 | 16.19 ± 0.04 | ... | ... | ... | ZTF |
| 2022-04-12 | 682.3 | 8.3 | ... | 15.36 ± 0.02 | ... | ... | ZTF |
| 2022-04-12 | 682.4 | 8.4 | 15.97 ± 0.06 | 15.37 ± 0.05 | 15.48 ± 0.07 | 15.56 ± 0.07 | GIT |
| 2022-04-14 | 684.2 | 10.2 | 16.34 ± 0.08 | ... | 15.54 ± 0.10 | 15.59 ± 0.09 | GIT |
| 2022-04-14 | 684.3 | 10.3 | ... | 15.45 ± 0.05 | ... | ... | GIT |
| 2022-04-17 | 687.4 | 13.4 | ... | 15.59 ± 0.03 | ... | ... | ZTF |
| 2022-04-18 | 688.2 | 14.2 | 16.84 ± 0.07 | 15.69 ± 0.07 | 15.63 ± 0.12 | 15.65 ± 0.09 | GIT |
| 2022-04-19 | 689.3 | 15.3 | ... | ... | ... | 15.76 ± 0.10 | GIT |
| 2022-04-19 | 689.4 | 15.4 | 16.91 ± 0.10 | 15.76 ± 0.06 | 15.71 ± 0.09 | ... | GIT |
| 2022-04-20 | 690.2 | 16.2 | ... | 15.84 ± 0.09 | ... | ... | GIT |
| 2022-04-20 | 690.3 | 16.3 | 17.09 ± 0.05 | ... | 15.73 ± 0.07 | 15.83 ± 0.15 | GIT |
| 2022-04-22 | 692.2 | 18.2 | 17.29 ± 0.06 | 15.99 ± 0.06 | 15.86 ± 0.09 | 15.86 ± 0.14 | GIT |
| 2022-04-23 | 693.2 | 19.2 | 17.42 ± 0.08 | ... | ... | 15.93 ± 0.11 | GIT |
| 2022-04-23 | 693.3 | 19.3 | ... | 16.07 ± 0.06 | ... | ... | GIT |
| 2022-04-24 | 693.5 | 19.5 | 17.44 ± 0.05 | ... | ... | ... | ZTF |
| 2022-04-24 | 694.2 | 20.2 | 17.39 ± 0.06 | 16.08 ± 0.07 | 15.96 ± 0.09 | 15.91 ± 0.11 | GIT |
| 2022-04-25 | 695.3 | 21.3 | 17.66 ± 0.09 | ... | ... | ... | ZTF |
| 2022-04-25 | 695.4 | 21.4 | ... | 16.02 ± 0.03 | ... | ... | ZTF |
| 2022-04-26 | 696.3 | 22.3 | 17.56 ± 0.06 | 16.23 ± 0.08 | 16.08 ± 0.08 | 15.98 ± 0.09 | GIT |
| 2022-04-27 | 697.3 | 23.3 | 17.69 ± 0.05 | 16.22 ± 0.11 | 16.16 ± 0.06 | 16.14 ± 0.09 | GIT |
| 2022-04-27 | 697.4 | 23.4 | ... | 16.29 ± 0.15 | ... | ... | ZTF |
| 2022-04-29 | 699.3 | 25.3 | ... | ... | ... | 16.13 ± 0.12 | GIT |
| 2022-04-29 | 699.4 | 25.4 | 17.78 ± 0.06 | 16.39 ± 0.08 | 16.26 ± 0.19 | ... | GIT |
| 2022-04-30 | 700.2 | 26.2 | 17.82 ± 0.05 | ... | ... | ... | ZTF |
| 2022-04-30 | 700.3 | 26.3 | ... | 16.37 ± 0.04 | ... | ... | ZTF |
| 2022-05-01 | 701.4 | 27.4 | ... | ... | 16.31 ± 0.20 | 16.26 ± 0.09 | GIT |
| 2022-05-02 | 702.3 | 28.3 | 17.83 ± 0.07 | ... | ... | ... | ZTF |
| 2022-05-02 | 702.4 | 28.4 | ... | 16.45 ± 0.05 | ... | ... | ZTF |
| 2022-05-05 | 705.4 | 31.4 | ... | 16.61 ± 0.12 | 16.42 ± 0.12 | ... | GIT |
| 2022-05-06 | 706.4 | 32.4 | 17.87 ± 0.10 | 16.67 ± 0.07 | 16.59 ± 0.18 | ... | GIT |
| 2022-05-07 | 707.2 | 33.2 | ... | 16.68 ± 0.03 | ... | ... | ZTF |
| 2022-05-07 | 707.3 | 33.3 | ... | 16.66 ± 0.04 | ... | ... | ZTF |
| 2022-05-08 | 708.3 | 34.3 | 17.94 ± 0.09 | 16.72 ± 0.07 | 16.57 ± 0.09 | 16.36 ± 0.09 | GIT |
| 2022-05-09 | 709.3 | 35.3 | ... | 16.77 ± 0.05 | ... | ... | ZTF |
| 2022-05-10 | 710.4 | 36.4 | ... | ... | ... | 16.56 ± 0.06 | GIT |
| 2022-05-11 | 711.2 | 37.2 | 18.07 ± 0.08 | ... | ... | ... | ZTF |
| 2022-05-11 | 711.3 | 37.3 | 18.13 ± 0.08 | 16.82 ± 0.05 | 16.86 ± 0.17 | 16.54 ± 0.10 | GIT |
| 2022-05-13 | 713.3 | 39.3 | 18.09 ± 0.07 | 16.86 ± 0.05 | ... | ... | ZTF |

**Table 6**
(Continued)

| Date | JD[a] | Phase[b] | $g$ | $r$ | $i$ | $z$ | Telescope |
|---|---|---|---|---|---|---|---|
| 2022-05-14 | 714.3 | 40.3 | 18.06 ± 0.11 | 16.94 ± 0.05 | 16.74 ± 0.12 | 16.59 ± 0.10 | GIT |
| 2022-05-15 | 715.3 | 41.3 | 18.03 ± 0.17 | 16.90 ± 0.06 | 16.68 ± 0.10 | 16.67 ± 0.11 | GIT |
| 2022-05-18 | 718.3 | 44.3 | ... | 17.05 ± 0.10 | 16.93 ± 0.09 | 16.84 ± 0.09 | GIT |
| 2022-05-19 | 719.2 | 45.2 | ... | 17.01 ± 0.04 | ... | ... | ZTF |
| 2022-05-19 | 719.3 | 45.3 | ... | ... | ... | 16.91 ± 0.08 | GIT |
| 2022-05-19 | 719.4 | 45.4 | 18.21 ± 0.09 | 17.04 ± 0.06 | ... | ... | GIT |
| 2022-05-21 | 721.3 | 47.3 | 18.16 ± 0.09 | 17.19 ± 0.10 | 16.98 ± 0.06 | 17.02 ± 0.08 | GIT |
| 2022-05-22 | 722.3 | 48.3 | 18.21 ± 0.05 | ... | ... | ... | ZTF |
| 2022-05-22 | 722.4 | 48.4 | 18.25 ± 0.08 | ... | ... | ... | ZTF |
| 2022-05-23 | 723.3 | 49.3 | 18.13 ± 0.06 | 17.17 ± 0.05 | 16.94 ± 0.11 | 16.93 ± 0.08 | GIT |
| 2022-05-24 | 724.3 | 50.3 | ... | 17.14 ± 0.05 | ... | ... | ZTF |
| 2022-05-24 | 724.4 | 50.4 | 18.23 ± 0.08 | ... | ... | ... | ZTF |
| 2022-05-25 | 725.3 | 51.3 | ... | ... | ... | 16.99 ± 0.10 | GIT |
| 2022-05-26 | 726.3 | 52.3 | 18.22 ± 0.11 | 17.16 ± 0.08 | 17.03 ± 0.12 | 17.08 ± 0.09 | GIT |
| 2022-05-26 | 726.4 | 52.4 | ... | 17.15 ± 0.05 | ... | ... | ZTF |
| 2022-05-29 | 729.3 | 55.3 | 18.34 ± 0.11 | 17.25 ± 0.12 | 17.16 ± 0.13 | 17.05 ± 0.08 | GIT |
| 2022-05-29 | 729.4 | 55.4 | 18.37 ± 0.08 | ... | ... | ... | ZTF |
| 2022-06-01 | 732.2 | 58.2 | ... | 17.33 ± 0.04 | ... | ... | ZTF |
| 2022-06-01 | 732.3 | 58.3 | 18.36 ± 0.06 | ... | ... | ... | ZTF |
| 2022-06-02 | 733.3 | 59.3 | 18.30 ± 0.09 | 17.40 ± 0.05 | 17.34 ± 0.06 | ... | GIT |
| 2022-06-03 | 734.2 | 60.2 | 18.32 ± 0.06 | 17.40 ± 0.05 | ... | ... | ZTF |
| 2022-06-04 | 735.3 | 61.3 | 18.33 ± 0.06 | 17.48 ± 0.05 | 17.33 ± 0.07 | 17.25 ± 0.07 | GIT |
| 2022-06-05 | 736.2 | 62.2 | ... | 17.48 ± 0.05 | ... | ... | ZTF |
| 2022-06-05 | 736.3 | 62.3 | 18.43 ± 0.09 | ... | ... | ... | ZTF |
| 2022-06-06 | 737.3 | 63.3 | 18.38 ± 0.09 | 17.50 ± 0.04 | 17.35 ± 0.05 | 17.25 ± 0.11 | GIT |
| 2022-06-07 | 738.2 | 64.2 | ... | 17.49 ± 0.07 | 17.35 ± 0.12 | 17.24 ± 0.08 | GIT |
| 2022-06-08 | 739.2 | 65.2 | 18.44 ± 0.09 | 17.55 ± 0.06 | 17.47 ± 0.06 | 17.26 ± 0.12 | GIT |
| 2022-06-09 | 740.2 | 66.2 | ... | 17.52 ± 0.06 | ... | ... | ZTF |
| 2022-06-09 | 740.4 | 66.4 | 18.47 ± 0.10 | ... | ... | ... | ZTF |
| 2022-06-11 | 742.2 | 68.2 | ... | 17.59 ± 0.05 | ... | ... | ZTF |
| 2022-06-11 | 742.3 | 68.3 | 18.45 ± 0.07 | 17.68 ± 0.04 | 17.50 ± 0.06 | 17.48 ± 0.11 | GIT |
| 2022-06-12 | 743.3 | 69.3 | ... | 17.71 ± 0.04 | ... | ... | GIT |
| 2022-06-16 | 747.2 | 73.2 | ... | 17.73 ± 0.04 | ... | ... | ZTF |
| 2022-06-16 | 747.3 | 73.3 | 18.59 ± 0.13 | ⋯ | ... | ... | ZTF |
| 2022-06-18 | 749.3 | 75.3 | 18.65 ± 0.08 | 17.75 ± 0.07 | ... | ... | ZTF |
| 2022-06-20 | 751.2 | 77.2 | 18.62 ± 0.08 | ⋯ | ... | ... | ZTF |
| 2022-06-20 | 751.3 | 77.3 | ... | 17.76 ± 0.06 | ... | ... | ZTF |
| 2022-06-22 | 753.2 | 79.2 | ... | 17.84 ± 0.05 | ... | ... | ZTF |
| 2022-06-22 | 753.3 | 79.3 | 18.58 ± 0.07 | 17.85 ± 0.04 | 17.65 ± 0.07 | ... | GIT |
| 2022-06-25 | 756.2 | 82.2 | ... | 17.92 ± 0.05 | ... | ... | ZTF |
| 2022-06-25 | 756.3 | 82.3 | 18.66 ± 0.08 | ... | ... | ... | ZTF |
| 2022-06-27 | 758.3 | 84.3 | ... | ... | ... | 17.82 ± 0.09 | GIT |
| 2022-06-30 | 761.2 | 87.2 | ... | 18.03 ± 0.06 | ... | ... | ZTF |
| 2022-07-11 | 772.1 | 98.1 | 19.08 ± 0.06 | 18.32 ± 0.11 | 17.92 ± 0.09 | 18.06 ± 0.09 | GIT |

**Notes.** Missing data are denoted by ellipses.
[a] JD offset = JD – 2459000.
[b] Phase in days relative to $g$-band maximum at JD = 2459674.00.

(This table is available in its entirety in machine-readable form in the online article.)

**Table 7**
Spectroscopic Observations of SN 2022eyw from HCT

| Date | JD[a] | Phase[b] (days) | Spectral Range (Å) |
|---|---|---|---|
| 2022-03-25 | 664.4 | −8.1 | 3700–7400 |
| 2022-03-26 | 665.4 | −7.1 | 3700–9100 |
| 2022-03-28 | 667.5 | −5.0 | 3700–7400 |
| 2022-03-29 | 668.5 | −4.0 | 5500–9100 |
| 2022-04-02 | 672.3 | −0.2 | 3700–9100 |
| 2022-04-03 | 673.2 | 0.6 | 3700–9100 |
| 2022-04-07 | 677.2 | 4.6 | 3700–9100 |
| 2022-05-03 | 703.4 | 30.5 | 3700–9100 |
| 2022-05-05 | 705.2 | 32.3 | 3700–9100 |
| 2022-05-08 | 708.3 | 35.4 | 3700–9100 |
| 2022-05-14 | 714.3 | 41.3 | 3700–9100 |
| 2022-05-18 | 718.2 | 45.2 | 3700–9100 |
| 2022-06-01 | 732.1 | 59.0 | 3700–9100 |
| 2022-06-13 | 744.3 | 71.0 | 3700–7580 |
| 2022-07-23 | 784.1 | 110.5 | 3700–7580 |

**Notes.**
[a] JD offset = JD − 2459000.
[b] All phases are given in rest-frame days since $B_{max}$ (= 2459672.55).

## ORCID iDs

Hrishav Das https://orcid.org/0009-0007-9727-7792
Devendra K. Sahu https://orcid.org/0000-0002-6688-0800
Anirban Dutta https://orcid.org/0000-0002-7708-3831
Mridweeka Singh https://orcid.org/0000-0001-6706-2749
G. C. Anupama https://orcid.org/0000-0003-3533-7183
Rishabh Singh Teja https://orcid.org/0000-0002-0525-0872